\documentclass[1p,preprint]{elsarticle}

\usepackage{amssymb}
\usepackage{amsmath}
\usepackage{lineno}
\usepackage{color}
\usepackage{listings}

\begin{document}

\begin{frontmatter}

\title{Solving large number of non-stiff, low-dimensional ordinary differential equation systems on GPUs and CPUs: performance comparisons of MPGOS, ODEINT and DifferentialEquations.jl}

\author[bme]{D\'aniel Nagy}
\ead{n.nagyd@gmail.com}

\author[bme]{Lambert Plavecz}
\ead{plaveczlambert@gmail.com}

\author[bme]{Ferenc Heged\H us\corref{cor1}}
\ead{fhegedus@hds.bme.hu}

\cortext[cor1]{Corresponding author}
\address[bme]{Department of Hydrodynamic Systems, Faculty of Mechanical Engineering, Budapest University of Technology and Economics, Budapest, Hungary}

\begin{abstract}
In this paper, the performance characteristics of different solution techniques and program packages to solve a large number of independent ordinary differential equation systems is examined. The employed hardware are an Intel Core i7-4820K CPU with 30.4 GFLOPS peak double-precision performance per cores and an Nvidia GeForce Titan Black GPU that has a total of 1707 GFLOPS peak double-precision performance. The tested systems (Lorenz equation, Keller--Miksis equation and a pressure relief valve model) are non-stiff and have low dimension. Thus, the performance of the codes are not limited by memory bandwidth, and Runge--Kutta type solvers are efficient and suitable choices. The tested program packages are MPGOS written in C++ and specialised only for GPUs; ODEINT implemented in C++, which supports execution on both CPUs and GPUs; finally, DifferentialEquations.jl written in Julia that also supports execution on both CPUs and GPUs. Using GPUs, the program package MPGOS is superior. For CPU computations, the ODEINT program package has the best performance.
\end{abstract}


\end{frontmatter}

\section{Introduction} \label{Sec:Introduction}

Low dimensional ordinary differential equation (ODE) systems are still widely used in physics and non-linear dynamical analysis. Among the simplest models (second or third-order systems), one can still find many publications employing the classic Duffng \cite{Luo2016a,Englisch2015a,Bonatto2008b,Englisch1991a,Gilmore1995a,Parlitz1985a}, Morse \cite{Krajnak2019a,Medeiros2013a,Knop1990a,Scheffczyk1991a}, Toda \cite{Goswami2011a,Goswami1998a,Goswami1996a,Kurz1988a}, Brusselator \cite{Deng2020a,Gallas2015a}, van der Pol \cite{Xu2019a,Mettin1993a,Parlitz1987a} and Lorenz \cite{Meucci2008a,Goswami2008a,Goswami2007a,Lorenz1963a} equations. From time to time, the results obtained in these simple test cases provide new insights into the complex dynamics of chaotic behaviour or into the bifurcation structure in multi-dimensional parameter spaces. There is an exhaustive literature of low dimensional systems in many other branches of physics; for instance, single bubble models in acoustic cavitation and sonochemistry \cite{Lauterborn2010a,Zhang2018a,Zhang2017a,Zhang2016a,Yasui2008a,Haghi2018a,Haghi2018b,Haghi2017a}, multi-species models in population dynamics \cite{Gyllenberg2019a,Cenci2018a,Zeeman1993a} or the problem of pressure relief valves \cite{Hos2015a,Hos2014a,Hos2012a,Hos2003a}, machine tools \cite{Molnar2017a,Altintas2008a} or wheel shimmy \cite{Takacs2009a,Takacs2008a} in engineering. Naturally, it is hard to give an exact definition for low dimension. In this paper, we follow a computational approach: a system is considered low dimensional if the application (solver) is not limited by memory bandwidth. Roughly speaking, this means that the entire problem (including the variables required by the solver) fits into the L1 cache of the CPU, or a large portion of the entire problem fits into the registers of the GPU. For the details of the different architectures, see Sec.\,\ref{Sec:CPUArchitectureProfiling} and Sec.\,\ref{Sec:GPUArchitectureProfiling}.

Although the precise description of the majority of the physical problems usually requires complex modelling (e.g., involving partial differential equations), the application of reduced-order models still play a significant role in many scientific fields. The obvious reason is that with increasing model complexity, the required computational capacity increases as well. Therefore, there is always a compromise between the size of the investigated system and the number of the parameter combinations that can be examined with a given computational capacity (within a reasonable time). Good examples are the computations of the escape rates in transient chaos \cite{Lai2010a} that need long transient simulations with millions of initial conditions, or parameter studies in high-dimensional parameter spaces, see a recent publication of a scan of nearly $2$ billion parameter combinations of an acoustic bubble \cite{Hegedus2020a}. In addition, in many cases, a low-order system is already an acceptable approximation of the physical process: the Keller--Miksis equation in bubble dynamics agrees very well with measurements \cite{Lauterborn2010a}, or moderately complex pressure relief valve models are able to predict the stability limit of the device \cite{Hos2015a}. \textit{The present paper focuses on the solution techniques of large parameter studies of such low-order systems.}

For CPUs, suits for solving ordinary differential equations have existed for decades. The codes of Harier \cite{Hairer1993a,Hairer1991a,Hairer_WebSite} and ODEPACK \cite{Hindmarsh1983a,ODEPACK_WebSite} written in Fortran are the ones having the longest history, and they are well-known packages in the community. The ODE suit SUNDIALS developed at the Lawrence Livermore National Laboratory (LLNL) also has a long history in the research of ODE solvers \cite{hindmarsh2005a,SUNDIALS_WebSite}. Besides the original Fortran 77 codes, the newest developments are done in ANSI C and C++. Another promising candidate for the solution of ODEs is ODEINT \cite{Ahnert2014a,ODEINT_WebSite} written in C++ to fully exploit its object-oriented capabilities. As final examples, all the major high-level programming languages, like Python (e.g., SciPy) \cite{Scipy_Integrate_WebSite}, Matlab \cite{Shampine1997a} or Julia (DifferentialEquations.jl) \cite{DifferentialEquations_jl_WebSite} provide a tool for ODEs. The interested reader is referred to one of the most exhaustive overviews of the available ODE suits in the market \cite{Rackauckas2018a}. It is expected that low-level, compiler-based solvers have excellent performance, while high-level languages are good for making clear and well-structured code, and excellent in fast and easy development. Some solutions (e.g., Julia) try to combine both and compile the code before execution (just-in-time compilation) providing a means for both fast and efficient code development. One of the many important questions is \textit{how effectively these program packages exploit the modern architectural features of CPUs for large parameter studies of low-dimensional ODEs; for instance, the SIMD capabilities via the vector registers.}

In contrast to CPUs, the general-purpose utilisation of GPUs for scientific computing is relatively new. It also needs a different way of thinking due to its massively parallel architecture. In case of large parameter studies where each instance of the ODE system is independent, the parallelisation strategy is simple; namely, a single GPU thread solves a single instance of an ODE system having different initial conditions or parameter sets. Even though the parallelisation strategy is trivial, many problems can still arise degrading the performance significantly. For example, GPU is a co-processor, and before an integration phase, data transfer via the extremely slow PCI-E bus is mandatory. If trajectory manipulations can be done only by the CPU between successive integration phases (e.g., the regular normalisation during the computation of the Lyapunov spectra), the performance can decrease. Unless, the program package provides the means to overlap CPU and GPU computations by asynchronous dispatch of GPU tasks (integration). Another example of an issue is the handling of events and non-smooth dynamical systems that can be a real challenge. For a detailed introduction to the possible issues, the interested reader is referred to our preliminary publication \cite{Hegedus2020b}.

Due to the aforementioned reasons, reimplementations of even the simplest Runge--Kutta solvers exist in the literature for many specialised problems: solving chemical kinetics \cite{Stone2018a,Niemeyer2014a,Stone2013a,Shi2012a}, simulation in astrophysics \cite{Brock2015a,Dindar2013a}, epidemiological model fitting \cite{Kovac2018a,AlOmari2013a} or non-linear dynamical analysis \cite{Fazanaro2016a,Rodriguez2015a}, to name a few. Some general-purpose solvers, e.g., ODEINT or DifferentialEquations.jl (written in Julia) offer the possibility to transfer the integration procedure to the GPU. However, the problem formulation must fit into a framework suitable for both CPU and GPU computations, which might result in suboptimal exploitation of the processing power of GPUs, as we shall see in Sec.\,\ref{Sec:KellerMiksisEquation} and Sec.\,\ref{Sec:ImpactDynamics}. To the best knowledge of the authors, only a few projects exist for developing a general-purpose solver tuned for GPU hardware. One is the program package MPGOS \cite{Hegedus2019a,Hegedus_MPGOS_WebSite,Hegedus_MPGOS_GitHub} written in C++ that has many built-in features: event handling, efficient trajectory manipulation during the integration, distribution of the workload to many GPUs and an easy way to overlap GPU and CPU computations. Its drawback is that only Runge--Kutta type solvers are available in its present version. Another package is ginSODA suitable for solving stiff ODE systems; however, according to its publication \cite{Nobile2018a}, it lacks the aforementioned special features of MPGOS.

The main aim of the present paper is to provide performance comparisons of three program packages by examining three different test cases using both CPUs and GPUs. In this way, the reader can easily select the best option for a given problem. Also, the introduced ideas can also help during the development of a new solver.

The first solver is MPGOS that is a natural choice as it is developed by Heged\H us, F., who is one of the co-authors of the present paper. It only supports integration on GPUs. Its main features have already been introduced in the previous section. Although MPGOS is written in C++, its interface is user-friendly, and the user needs no knowledge about GPU programming for good performance. The second choice is ODEINT written in C++, that is a general-purpose solver having many features including different types of solvers. However, it is hard to use and not user-friendly. A good performance is expected at least for CPUs. Using the appropriate data structure (e.g., Thrust) integration can also be transferred to the GPU. ODEINT is part of the boost library collection \cite{Boost_WebSite}. The third candidate is the program package DifferentialEquations.jl written in Julia and capable of performing integration on both CPUs and GPUs. It has many built-in features: stiff problems via automatic differentiation, delay differential and algebraic differential equations, stochastic differential equations, symplectic solvers, boundary value problems and many more. Julia is a high-level language offering an easy-to-use development environment like Matlab or Python, but it compiles the code before the first run. Therefore, good performance is expected combined with user-friendliness. In the present paper, only the above-described three solvers are tested to remain focused and to avoid an overwhelming flow of data.

Altogether three systems are examined. The first test case performs $1000$ steps with the classic fourth-order Runge--Kutta scheme with fixed time steps on the classic Lorenz system \cite{Lorenz1963a}. This is a standard test case provided in the majority of the program packages as tutorial or reference example. The amount of work that needs to be done is well defined as it does not depend on the initial conditions or on the parameter sets (fixed time steps, no error control). Thus, there is no thread divergence using GPUs. Second, an amplification diagram is simulated employing the Keller--Miksis equation \cite{Lauterborn2010a,Keller1980a} describing the radial pulsation of a spherical gas bubble placed in an infinite domain of liquid. This model is a perfect example to test the codes when even a single trajectory has orders of magnitude differences in the time scale during an integration process. On GPUs, this feature of the system can result in a large amount of thread divergence, or a tremendous amount of overhead of the computation in case the time stepping cannot be separated between the systems. The third model describes the dynamics of a pressure relief valve \cite{Hos2012a} that can exhibit impact dynamics. In this example, the performance of the packages can be tested when multiple events have to be detected. The events occur at different time instances corresponding to the different systems. Moreover, upon the detection of an impact as an event, the impact law has to be applied immediately on the specific system. It must be stressed that the impacts are not ``synchronised'', each system has its own unique history of impacts that makes this problem a challenge for GPU codes.

\section{Limiting factors of a computation} \label{Sec:LimitingFactors}

The performance of an application running on either CPU, GPU or both can be limited by memory bandwidth, by saturation of the compute units or by latency. In case of memory bandwidth limited applications, the bandwidth (usually measured in Gb/s) of the global memory (GPU) or system memory (CPU) bus system is not enough to feed the processing units with enough data. Thus, these units are idle in some portion of the total runtime. A perfect example is a matrix-matrix multiplication with matrices so large that they do not even fit into the last level (L3) cache of the CPU or shared memory of the GPU. In this case, with a na\"ive approach, the elements of an operation have to be loaded from the global or system memory (again) even if that particular data was already loaded before. The reason is the limited amount of cache: some old data need to be evicted in order to load another portion. The optimal number of load operations is $2 \times N^2$ (every element loaded only once) in contrast to the worst-case scenario: $2 \times N^4$ (all elements are loaded again). Here, $N \times N$ is the size of the matrices. The difference between the optimal and the worst-case scenarios is huge since $N$ is large. It can be calculated by paper and pencil that the memory bandwidth required in the worst-case scenario is usually orders of magnitude larger than the available. Many smart approaches exist to ease the pressure on the global or system memory by exploiting the memory hierarchy of the hardware. For instance, performing as much calculation as possible on submatrices loaded into the L1 cache of the CPU or into the shared memory of the GPU. This follows \textit{the concept of data reuse}.

\textit{For low dimensional ODE systems (our case), the available memory bandwidth is usually not an issue.} A large portion of the data fits into the registers, or the L1 cache or shared memory. Memory bandwidth plays an important role for very large problems when the ODE system comes, e.g., from the semi-discretisation of one or more partial differential equations. In this case, usually many matrix-vector multiplications are involved during the evaluation of the right-hand side of the ODE function.

The second type of limiting factor of an application is when the compute units of the hardware are fully utilised. This is a desirable operation condition as the main aim is to harness the peak processing power of the hardware. It must be stressed, however, that a compute limited application is not necessarily efficient. It is possible that a wrong choice of a numerical scheme needs orders of magnitude larger number of floating-point operations compared to an optimal algorithm.

If both the utilisation of the memory bus system and the arithmetic processing units are low (the code is neither limited by memory bandwidth nor by the utilisation of the compute units), the limiting factor is the latency. Performing an operation, the results are available only after a certain amount of time (measured in clock cycles). This ``delay'' is called latency. Keep in mind that latency has nothing to do with memory bandwidth (peak data transfer in Gb/s) or arithmetic throughput (peak processing power in FLOPS). Even in case of a request of a single floating-point number from the global memory (the memory bandwidth is definitely not saturated), the data is available for computations only after approximately $600$ clock cycles. Similarly, even if an addition or a multiplication can be performed within a single clock cycle, the computed results are available for further use only after $3-5$ clock cycles (architecture and instruction dependent). The solution to this issue is to increase parallelism. In the case of CPUs, it can be done by breaking long dependency chains in the code so that the CPU core has more options to select a non-dependent operation. A common example is the loop unrolling technique where the statements inside the loop are independent of each other and can be computed in parallel. Practically, in our case, this means the solution of more ODE systems together at the same time even on a single CPU core. For GPUs, the latency can be hidden mainly by launching enough computing threads during kernel execution. Thus, the GPU has more ``chance'' to select threads ready for computation. A comprehensive overview of the differences between CPU and GPU architectures and computational strategies are introduced in the textbook \cite{Soyata2018a}.

\section{An introductory example} \label{Sec:IntroductoryExample}

In this section, the techniques to harness the peak processing power of CPUs and GPUs are demonstrated via the simple one-dimensional ODE system written as
\begin{equation} \label{exemplary_system}
\dot{x} = x^2 - p,
\end{equation}
where $x$ is the sole state variable and $p$ is a parameter. Since this paper focuses on extensive parameter studies, many instances of Eq.\,\eqref{exemplary_system} are solved each having a different parameter value. The dot stands for the derivative with respect to time. The numerical algorithm is the classic fourth-order Runge--Kutta scheme with fixed time steps. Thus, the required number of arithmetic operations is independent of the initial condition and the parameter value. The reason for the choice of Eq.\,\eqref{exemplary_system} is the structure of its right-hand side; namely, it consists of exactly one addition and one multiplication. In modern CPU and GPU architectures, these two operations can be performed via a single fused multiply-add (FMA) instruction. It is important as the peak theoretical processing power of the CPUs or GPUs are computed with the assumption that every floating-point instructions are FMAs. Consequently, a numerical problem consisting only of non-FMA instructions (e.g., sole additions and multiplications) can harness only half the theoretical processing power even in the best-case scenario. Since the Runge--Kutta solver itself includes some non-FMA instructions, $100\%$ exploitation of the peak processing power cannot be expected.

\subsection{Maximising the performance on CPU} \label{Sec:MaximisingPerformanceCPU}

\subsubsection{The na\"ive approach} \label{Sec:NaiveApproach}

The simplest (na\"ive) approach to solve a large ensemble of Eq.\,\eqref{exemplary_system} is simply to loop through the parameter values and perform the integration one after another. The corresponding simplified code snippet is listed in Lst.\,\ref{NaiveApproach}. The control parameter $p$ is varied between $0.1$ and $1.0$ with a resolution of $N=2^{16}=65536$ on an equidistant grid. Altogether $n=1000$ fourth-order Runge--Kutta steps are computed. The dense output is not stored; thus, a single Runge--Kutta step updates the state variable $x$. As the system is autonomous, it is unnecessary to register the actual time $t$. The final time instances can be obtained by $T=n \cdot dt$.

\begin{lstlisting}[float, language=C++, basicstyle=\small\selectfont\ttfamily, commentstyle=\color{red}, keywordstyle=\color{blue}\bfseries, label=NaiveApproach, caption=Simplified code snipped of the na\"ive approach for solving Eq.\,\eqref{exemplary_system}.]
const int N = 1<<16;
const int n = 1000;
const double dt = 0.01;

double* P = linspace(0.1, 1.0, N);
double x;
double p;

for (int i=0; i<N; i++)
{
    p = P[i];
    x = -0.5;
    
    for (int j=0; j<n; j++)
        OneStep(x,p,dt);
}
\end{lstlisting}

The runtime of the code was $2.383\,\mathrm{s}$ on an Intel Core i7-4820K CPU (Ivy Bridge architecture) using a single core. Although the runtime is an important measure of code performance, it does not provide any information on how much of the peak theoretical performance of the CPU core is harnessed. For this, a specialised profiling technique has to be used in order to monitor the total number of floating-point instructions, see Sec.\,\ref{Sec:CPUArchitectureProfiling} for more details. For code snippet Lst.\,\ref{NaiveApproach}, it turned out that the achieved GFLOPS (Giga floating-point operations per second) is $1.155$, whereas the peak theoretical power of the core is $30.4\,\mathrm{GFLOPS}$. This means a mere $3.9\%$ FLOPS efficiency. This value is surprisingly low and to improve the code, a deeper understanding of the hardware architecture and the profiling technique is required, see Sec.\,\ref{Sec:CPUArchitectureProfiling}.

\subsubsection{Basics of CPU architectures and profiling} \label{Sec:CPUArchitectureProfiling}

\begin{figure}[ht!]    
	\centering
		\includegraphics[width=8.6cm]{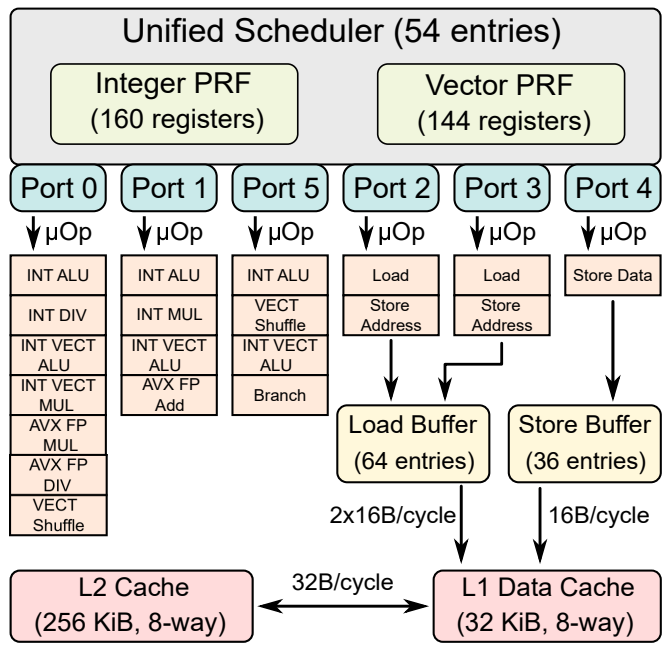}
	\caption{The compute engine (back end) of the Intel Ivy Bridge micro-architecture.}
	\label{Fig:IvyBridgeBackEnd}
\end{figure}

The basics of the CPU architectures are introduced via the description of the compute engine (back end) of the Intel Ivy Bridge micro-architecture, see Fig.\,\ref{Fig:IvyBridgeBackEnd}. This is the architecture of the CPU used throughout this study. In spite of this specific example, the main conclusions hold for other types of processors as well. In Fig.\,\ref{Fig:IvyBridgeBackEnd}, the front end responsible for the decoding of the instructions into micro-operations are omitted as usually it is not the bottleneck of the solution of ODEs. The execution engine of Ivy Bridge (and every modern CPUs) executes micro-operations in an out-of-order fashion. That is, the instructions are not necessarily executed exactly in the order as they are in the assembly code. Any micro-operation can be dispatched to one of the six ports for execution if it does not depend on the results of other micro-operations. In this sense, a maximum of six micro-operations can be dispatched for execution at the same time. This represents the instruction-level parallelism (ILP) of the CPU. Note how the different ports are responsible for the execution of different types of instructions. For example, ports $0$, $1$ and $5$ are responsible for arithmetic computations, while ports $2-4$ are responsible for memory management (load/store and address calculation). Therefore, floating-point calculations and load requests for subsequent computations can be overlapped.

From Fig.\,\ref{Fig:IvyBridgeBackEnd}, it is clear that a floating-point multiplication (port $0$) and a floating-point addition (port $1$) can be performed simultaneously. This alone implies that a certain amount of ILP must be presented in the code. For the techniques of increasing ILP, the reader is referred to Sec.\,\ref{Sec:InstructionLevelParallelism}. The instruction throughput of floating-point addition/subtraction and multiplication is $1$ per clock cycle. However, the latency of multiplication is $5$ cycles, and the latency of addition/subtraction is $3$ cycles. Therefore, an even larger ``amount'' of ILP must be presented in the code to be able to feed ports $0$ and $1$ with enough instructions. The issue of fully hiding the latency is problem-dependent, one has to make experiments system-by-system. Observe that the Ivy Bridge microarchitecture does not support the fused multiply-add (FMA) instructions. This feature, however, is already built-in in the next-generation Haswell processors. Nevertheless, the equal number of addition/subtraction and multiplication is a must to get close to the peak theoretical performance.

Another important issue one has to consider is the division operation dispatched to port $0$. It has a reciprocal throughput of $20-44$ clock cycles per instructions, and latency of $21-45$ cycles. That is, it is a very expensive operation. In addition, it blocks port $0$ for multiplications further reducing the performance of the code. Therefore, if a division of a number is required several times, compute its reciprocal in a separate variable and use it as a multiplicator. The negative effect of including division is demonstrated in Sec.\,\ref{Sec:SpecialFunctionsDivisions}. For a thorough analysis of the throughput and latency of different instructions on different microprocessor architectures (Intel, AMD and VIA), the interested reader is referred to the excellent work of Fog, A. \cite{Fog2020a}.

Observe the AVX prefix in front of the floating-point addition, multiplication and division panels in Fig.\,\ref{Fig:IvyBridgeBackEnd}. This means the support of the Advanced Vector Extension instruction set to perform Single Instruction Multiple Data (SIMD) operations. That is, the same operation is performed on different pieces of data. This instruction set uses vector register (YMM) that are capable of holding $4$ double or $8$ single-precision floating-point numbers (that is, a YMM register is $256$ bit wide). Thus, an application using doubles can experience a $4$ times speed-up if the vector capabilities of the CPU can be fully exploited. Conversely, an application cannot even get close to the theoretical peak performance without the use of the AVX instruction set. In Sec.\,\ref{Sec:ExplicitVectorisation}, a very easy technique is introduced to ensure the usage of vector registers for parameter studies of ODE systems. The AVX instruction set was introduced in the Sandy Bridge (Intel) and Bulldozer (AMD) processors. The AVX2 instruction set is available since CPU generations Haswell (Intel) and Excavator (AMD). Its main novelty is the support of fused multiply-add (FMA) instructions doubling the peak theoretical performance compared to AVX. The latest version of the Advanced Vector Extension set is AVX-512 introduced in the microarchitecture Skylake (Intel). It is capable of performing instructions on $8$ doubles or $16$ floats further doubling the peak theoretical performance.

In order to get a detailed picture about the behaviour of the CPU core, the \textbf{perf} profiler utility \cite{Perf_WebSite} was used, which is a tool for Linux based systems. It can collect events and metrics from the Performance Monitoring Unit (PMU) of the processor such as the number of cycles, instructions retired, L1 cache misses or instructions issued to each port, to name a few. Although the achieved floating-point performance and efficiency cannot be retrieved directly, they can be calculated by the elapsed time and the number of the executed floating-point instructions. A typical output for a few events are presented in Lst.\,\ref{OutputPerf}. The discussion of the full capabilities of \textbf{perf} is beyond the scope of the present paper. As a tabulated data, the most important events and indirectly calculated data (e.g., achieved GFLOPS and FLOP efficiency) are summarised in Tab.\,\ref{ProfIntroductoryExample}. The peak processing power of the employed Intel Core i7-4820K CPU is 30.4 GFLOPS (single core). The operating system is Ubuntu 20 LTS, and the C++ compiler is gcc 7.5.0. The column with the header \textit{simple} is related to the na\"ive approach introduced in Sec.\,\ref{Sec:NaiveApproach}. It can be seen that the number of clock cycles of the ``stall reason due to no eligible RS entry is available'' (``No RS'' in the table for short) is only slightly smaller than the total number of clock cycles. The abbreviation RS is the Reservation Station, also called the Unified Scheduler, see Fig.\,\ref{Fig:IvyBridgeBackEnd}. This implies that the CPU is idle most of the time because data is not available from a previous computation (most likely) or it has not arrived from the memory subsystem (less likely). This explains the poor performance and low FLOPS efficiency. Note that Tab.\,\ref{ProfIntroductoryExample} collects profiling data also for all the other versions of the introductory example Eq.\,\eqref{exemplary_system} discussed in the following subsections, and for the test cases of using divisions and transcendental functions, see Sec.\,\ref{Sec:SpecialFunctionsDivisions}.

\begin{lstlisting}[float, basicstyle=\small\selectfont\ttfamily, label=OutputPerf, caption=Typical output of a profiling with the utility \textbf{perf}.]
10.049.141.981   Number of clock cycles
18.775.076.456   Number of instruction
 9.759.952.395   Counts 256-bit packed double-precision
 6.579.729.696   L1-dcache-loads
       119.756   L1-dcache-load-misses
 4.252.220.453   L1-dcache-stores
        57.138   L1-dcache-store-misses
 4.698.542.488   Cycles in which a uop is dispatched on port 0
 5.065.437.974   Cycles in which a uop is dispatched on port 1
 5.264.870.995   Cycles in which a uop is dispatched on port 2
 5.568.711.371   Cycles in which a uop is dispatched on port 3
 4.261.921.012   Cycles in which a uop is dispatched on port 4
 1.979.687.970   Cycles in which a uop is dispatched on port 5
 6.955.451.533   Cycles stalled due to Resource Related reason
 2.684.174.187   Cycles stalled due to no elig. RS entry avail.
 4.639.678.096   Cycles stalled due to no store buffers avail.
         7.148   Cycles stalled due to re-order buffer full
\end{lstlisting}

\begin{table}
  \caption{Summary of the condensed representation of the profiling information for the introductory examples. The peak processing power of the employed Intel Core i7-4820K CPU is 30.4 GFLOPS (single core). The operating system is Ubuntu 20 LTS, and the C++ compiler is gcc 7.5.0.}
  \label{ProfIntroductoryExample}
  \begin{equation*}
    \begin{array}{|c|ccccc|}
      \hline
     \text{Code} & \text{simple} & \text{simple} & \text{simple} & \text{transc.} & \text{division} \\
     \text{} & \text{vcl unroll} & \text{vcl} & \text{--} & \text{vcl unroll} & \text{vcl unroll} \\
     \hline \hline
     \text{Unroll} & 8 & 1 & 1 & 2 & 2 \\
     \text{Runtime [s]} & 0.1342 & 0.5975 & 2.383 & 2.125 & 1.004 \\
     \text{Dev [s]} & 0.0002 & 0.0010 & 0.0047 & 0.1645 & 0.0018 \\
     \text{Clock Cycles} & 4.960\cdot 10^8 & 2.209\cdot 10^9 & 8.811\cdot 10^9 & 7.643\cdot 10^9 & 3.711\cdot 10^9 \\
     \text{Number of Instr.} & 9.193\cdot 10^8 & 7.602\cdot 10^8 & 3.025\cdot 10^9 & 1.277\cdot 10^{10} & 8.587\cdot 10^8 \\
     \text{Number of X87} & 10480 & 17180 & 41680 & 87620 & 72620 \\
     \text{Number of SSE} & 1.319 \cdot 10^5 & 1.319\cdot 10^5 & 2.753\cdot 10^9 & 5.247\cdot 10^8 & 132100. \\
     \text{Number of AVX} & 7.088\cdot 10^8 & 6.882\cdot 10^8 & 0 & 5.736\cdot 10^9 & 8.197\cdot 10^8 \\
     \text{GFLOPS} & 21.12 & 4.607 & 1.155 & 11.04 & 3.267 \\
     \text{Efficiency [$\%$]} & 71.36 & 15.56 & 3.904 & 37.31 & 11.04 \\
     \text{L1 Cache Loads} & 3.413\cdot 10^8 & 1.510\cdot 10^6 & 2.235\cdot 10^6 & 5.097\cdot 10^9 & 3.434\cdot 10^7 \\
     \text{L1 Cache Misses} & 9.378 \cdot 10^4 & 1.128 \cdot 10^5 & 1.839 \cdot 10^5 & 2.055 \cdot 10^5 & 1.280 \cdot 10^5 \\
     \text{Divider Active} & 5.082 \cdot 10^4  & 8.369 \cdot 10^4 & 2.004 \cdot 10^5 & 4.067 \cdot 10^5 & 3.687\cdot 10^9 \\
     \text{Divider [$\%$]} & 0.01025 & 0.0037 & 0.0022 & 0.0053 & 99.35 \\
     \text{No store buffer} & 1.185\cdot 10^6 & 391500. & 545400. & 519000. & 419700. \\
     \text{No RS} & 2.581\cdot 10^8 & 1.996\cdot 10^9 & 8.018\cdot 10^9 & 4.060\cdot 10^9 & 3.410\cdot 10^9 \\ \hline
    \end{array}
  \end{equation*}
\end{table}

\subsubsection{Explicit vectorisation} \label{Sec:ExplicitVectorisation}

In Sec.\,\ref{Sec:CPUArchitectureProfiling}, it has been shown that vectorisation (using the vector registers) is mandatory for good performance. An optimising compiler might automatically vectorise the code; for instance, by unrolling a loop by a factor of two, four or eight depending on the available instruction set. The unrolling can be done if the loop body can be executed independently. This is exactly the case for the outer cycle in Lst.\,\ref{NaiveApproach}. However, due to the inner loop, the relatively complex structure of a single step and the call to an ``external'' ODE function, the compiler cannot recognise that cycling through the parameters are independent and can be done in parallel in a vectorised form. Therefore, in most of the cases, the best practice is to make the vectorisation by ourselves. We call this explicit vectorisation.

One way to implement explicit vectorisation is to use the intrinsic functions of the instruction set directly. It is a cumbersome task as even a simple addition has to be done via the function call  
\begin{verbatim}
__m256d _mm256_add_pd (__m256d a, __m256d b)
\end{verbatim}
for double-precision floating-point numbers. This makes the code hard to read. In addition, there are no intrinsics for trigonometric and other special functions like exponentials, logarithms or powers. Naturally, this difficulty can be overcome with operator and function overloading (this also needs the reimplementation of the special functions). For C++, there are libraries that offer full support of explicit vectorisation. Throughout this paper, the Vector Class Library (VCL) written by Fog, A. \cite{Fog2020b,Fog_VCL_GitHub} is applied.

\begin{lstlisting}[float, language=C++, basicstyle=\small\selectfont\ttfamily, commentstyle=\color{red}, keywordstyle=\color{blue}\bfseries, label=ExplicitVectorisation, caption=Simplified code snippet of the explicit vectorisation approach for solving Eq.\,\eqref{exemplary_system}.]
#include "vectorclass.h"
...
const int N = 1<<16;
const int n = 1000;
const double dt = 0.01;

double* P = linspace(0.1, 1.0, N);
Vec4d x;
Vec4d p;

for (int i=0; i<N; i+=4)
{
    p.load_a(P + i); // Loads parameters from aligned memory
    x = -0.5;
    
    for (int j=0; j<n; j++)
        OneStep(x,p,dt);
}
\end{lstlisting}

Listing.\,\ref{ExplicitVectorisation} demonstrates that with the VCL library, only minor modifications are necessary to exploit the vectorisation capabilities of a CPU. Besides the inclusion of the related header file, only the data types are replaced from double to Vec4d (packed $4$ doubles), a special function is used to load $4$ parameters from a memory location, and the counter of the outer loop is incremented by $4$. Naturally, inside the stepper and the ODE function, the data types have to be replaced as well (not shown here).

The profiling results of the new code are shown in Tab.\,\ref{ProfIntroductoryExample} in the column with the header \textit{simple vcl}. There is a massive improvement ($\times 3.988$) in the runtime indicating the full exploitation of the capabilities of the vector registers. Accordingly, the calculated FLOPS efficiency is also increased approximately by a factor of $4$; however, it is still only $15.56\%$ of the theoretical peak performance. As the stall reason due to no RS entry (last row in Tab.\,\ref{ProfIntroductoryExample}) is still very high, the code is still bound by latency. Since the number of the L1 cache loads are very low, the latency is caused by the dependency of the arithmetic operations.

\subsubsection{Increase instruction-level parallelism (ILP)} \label{Sec:InstructionLevelParallelism}

Due to the reasons discussed in Sec.\,\ref{Sec:CPUArchitectureProfiling} (out-of-order execution, the latency of the instructions), the increase of instruction-level parallelism is usually necessary to harness the full potential of the CPU cores. Solving ODE systems, two main approaches to increase ILP are discussed in this section. One can pack multiple independent ODE systems into a single right-hand side function as follows
\begin{align} \label{array_exemplary_system}
\begin{split}
\dot{x}_1 &= x_1^2 - p_1, \\
\dot{x}_2 &= x_2^2 - p_2, \\
\cdots \\
\dot{x}_m &= x_m^2 - p_m.
\end{split}
\end{align}
That is, altogether $m$ one-dimensional ODE systems described by Eq.\,\eqref{exemplary_system} are solved simultaneously. As each line in Eq.\,\eqref{array_exemplary_system} is independent, they can be evaluated in an out-of-order fashion. Naturally, this approach can be combined with vectorisation where each line in Eq.\,\eqref{array_exemplary_system} represents a pack of $2$, $4$ or $8$ systems (depending on the size of the vector registers). It means an unroll of $2m$, $4m$ or $8m$ of the parameters. In this source of ILP, the full-pack of systems has common time steps. For some problems, it is a severe issue responsible for the orders of magnitude increase in runtime, for details see Sec.\,\ref{Sec:KellerMiksisEquation}. However, in the cases of ODEINT and DifferentialEquations.jl, this is the only way the ILP can be increased as the solvers are wrapped into the program packages (unless the users implement their own solver). The advantage of this approach is its simplicity, and that the solver algorithm does not need to be rewritten or modified. This approach works fine with solvers using fixed time steps or with systems where the required time steps to keep a prescribed tolerance do not change significantly.

\begin{lstlisting}[float, language=C++, basicstyle=\small\selectfont\ttfamily, commentstyle=\color{red}, keywordstyle=\color{blue}\bfseries, label=InstructionLevelParallelism, caption=Simplified code snippet of the instruction level parallelism approach for solving Eq.\,\eqref{exemplary_system}.]
#include "vectorclass.h"
...
const int N = 1<<17;
const int n = 1000;
const int m = 8;
const double dt   = 0.01;
const double dtp2 = dt/2;

double* P = linspace(0.1, 1.0, N);
Vec4d x[m];
Vec4d p[m];
Vec4d kSum[m];
Vec4d kAct[m];

const int PStep = 4*m;
for (int i=0; i<N; i+=PStep)
{
    for (int j=0, offset=i; j<m; j++, offset+=4)
    {
        p[j].load_a(P+offset); // Load parameters
        x[j] = -0.5;
    }
    
    for (int j=0; j<n; j++)
    {
        // Unrolled automatically by the compiler
        for (int k=0; k<m; k++) // k1
        {
            F(&(kAct[k]), x[k], p[k]); // k1 = F(x)
            kSum[k] = kAct[k];
            kAct[k] = x[k] + dtp2*kAct[k];
        }
        ...
    }
}
\end{lstlisting}

For our ``hand-tuned'' C++ codes, a different approach is applied where the time coordinates can be kept separated, and each pack of systems (packed only via the VCL library) can be solved ``asynchronously''. The idea is to transfer the unroll procedure inside the solver and do it at a stage-level of the Runge--Kutta algorithm. The extended code snippet for this approach is presented in Lst.\,\ref{InstructionLevelParallelism}, where only the first stage of the full code is shown for brevity. The compiler can unroll the innermost loop automatically, as its structure is simple enough, with a constant unroll factor $m$ (known at compile time). Thus, several independent code blocks are created, increasing the available ILP, which can easily be regulated by the unroll factor $m$. Note that the variables $x$ and $p$ are arrays of size $m$ and of type Vec4d (packed $4$ doubles). Accordingly, the outermost loop responsible for cycling through the parameters has to be incremented by $4 m$ that is also the number of the simultaneously solved ODE systems.

Observe that for cases with fixed time steps, there is no real difference between the two strategies (unroll at ODE function level, and unroll at Runge--Kutta stage level). However, the latter one allows the use of different time steps by defining a vector as follows
\begin{verbatim}
const Vec4d dt[m];
\end{verbatim}
and apply different time steps for different instances of Eq.\eqref{exemplary_system} at the Runge--Kutta stages of an adaptive solver. In this way, all instances are solved asynchronously. As the solver shown in Lst.\,\ref{InstructionLevelParallelism} is the classic fourth-order Runge--Kutta scheme, the implementation of the application of different time steps unnecessarily increases the overhead without a real benefit (thus it is excluded). The main message of this section is still clear: in one way or another, the ILP has to be increased for latency-limited applications.

The profiling results for the optimal unroll factor ($m=8$) are presented in Tab.\,\ref{ProfIntroductoryExample} at the column with the header \textit{simple vcl unroll}. Again, a massive performance increase can be observed. The code is now close to the theoretical peak processing power (up to $71\%$). The stall reason due to no RS entry (last row in Tab.\,\ref{ProfIntroductoryExample}) is dropped by almost an order of magnitude. Interestingly, the L1 cache loads increased drastically implying that the larger number of ODE systems cannot fit into the physical registers of the core. That is, not only the dependency of the arithmetic instructions are responsible for the latency but also the data requests from the L1 cache. For more detailed information on the effect of the unroll factor $m$ (up to $m=16$), the interested reader is referred to the GitHub repository of the problem \cite{Basic_tests_RK4_GitHub}. Proposing further code optimisation strategies are beyond the scope of the present paper. With the explicit vectorisation and the unroll techniques, a large portion of the peak performance of the CPU can be harnessed.

\subsection{Maximising the performance on GPU} \label{Sec:MaximisingPerformanceGPU}

\subsubsection{Basics of GPU architectures and profiling} \label{Sec:GPUArchitectureProfiling}

The basic logical unit performing calculations is a thread. The number of threads simultaneously residing in a GPU can be in the order of hundreds of millions. This is the reason for the widely used term: massively parallel programming. In general, threads in a GPU are organised in a 3D structure called a grid. For our purpose, a 1D organisation is sufficient. That is, a unique identifier of a thread can be characterised by a 1D integer coordinate. The parallelisation technique to solve an ensemble of ODE systems is simple: a single instance of the ODE system (e.g., Eq.\,\eqref{exemplary_system}) is assigned to a single GPU thread. That is, the different threads work on different sets of data (parameters or initial conditions). This is the well-known per-thread approach \cite{Stone2018a}.

\begin{figure}[ht!]    
	\centering
		\includegraphics[width=12.5cm]{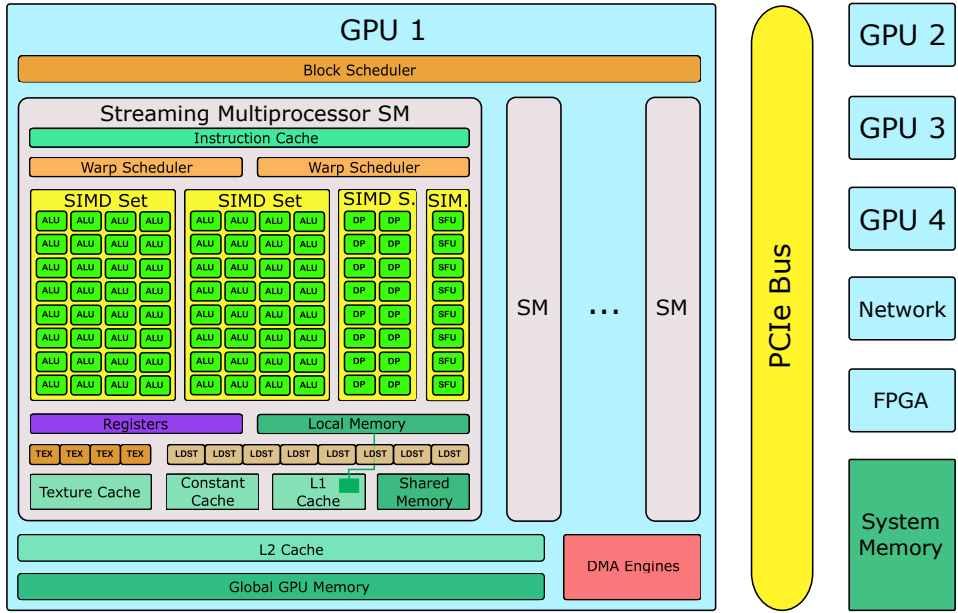}
	\caption{A general block diagram of GPU architectures. The details (e.g., the number of warp schedulers and compute units per streaming multiprocessor) are revision-specific features of the architectures. Thus they can vary from GPU to GPU.}
	\label{Fig:GPUArchitectureGeneral}
\end{figure}

Each GPU consists of one or more streaming multiprocessors (SMs) which are at the highest level of hierarchy in the hardware compute architecture. Each SM contains many processing units capable of performing integer or single-precision floating-point operations (ALU), double-precision floating-point operation (DP), load/store operations from the global memory or from the L1 cache/shared memory (user-programmable cache) or other specialised instructions, see Fig.\,\ref{Fig:GPUArchitectureGeneral}. The logical threads have to be mapped to the GPU architecture. For this purpose, the threads are partitioned into thread blocks. The workload in a GPU is distributed to the streaming multiprocessors (SMs) with block granularity. That is, the block scheduler of the GPU fills every SM with blocks of threads until reaching hardware or memory resource limitations.

The thread blocks are further divided into smaller chunks of execution units called warps. Each warp contains $32$ threads (as a current CUDA architectural design). The warp schedulers of an SM take warps and assign them to execution units if there are eligible warps for executions and there are free execution units on the SM. A warp is eligible if there is no data dependency from another computational phase or all the required data has already arrived from a memory load operation. Every thread in a warp performs the same instruction but on different data. This technique is known as single instruction multiple data paradigm (SIMD). Therefore, every thread in a warp executes their instruction stream in lock-step. This is the only way the hardware can efficiently handle a massive number of threads: for $32$ threads, only one control unit is necessary to track their program state. This results in more place for compute units and higher arithmetic throughput. If an SIMD set in the SM has not enough compute units (e.g., the number of the DP units are $16$ in an SM in Fig.\,\ref{Fig:GPUArchitectureGeneral}), the whole warp is executed in multiple launches (in the case of DP unit, only half of the warp is executed at the same time).

The drawback of the SIMT approach is the possibility of thread divergence occurring when some threads have to do different instructions from the others in the same warp. The simplest case is when the number of the required time steps are different for each system (thread) in an adaptive algorithm. The total execution time of a warp will be determined by its slowest thread forcing the others to be idle. \textit{To minimize the effect of thread divergence, the consecutive systems should have similar parameters and/or initial conditions to ensure similar dynamical behaviour in a warp.}

It is common in both CPUs and GPUs that the system or global memory cannot feed the compute units (in general) with enough data to utilise them fully. Therefore, each architecture heavily relies on its memory subsystem to ease the pressure on the slow system or global memories. The register memory and cache hierarchy are the main components of the memory subsystem. Registers are the fastest memory types (practically without latency); every instruction can be performed only on data already residing in the registers. The cache hierarchy is a tool to prefetch data to fast, low latency memory types and keep the data there for frequent reuse (if possible). Going higher in the cache hierarchy---from L3 (only in CPUs) to L2 and L1---their sizes but also the latencies decrease.

The main strategies of CPUs and GPUs for hiding latency are different. This includes latency by computational data dependency (approximately $3-40$ cycles depending on the instruction, see the discussion in Sec.\,\ref{Sec:CPUArchitectureProfiling}), although the latency of system or global memory reads/writes is much higher (approximately $600$ cycles or more). CPUs follow the latency-oriented design; that is, a single thread (or two in case of hyper-threading) in a core uses a small number of registers but a large amount of low-latency L1 and L2 caches (the last level L3 cache is shared among the cores). For instance, in the Ivy Bridge architecture shown in Fig.\,\ref{Fig:IvyBridgeBackEnd}, the physical register file size is $144$ entries of $256$ bit wide YMM registers (according to the AVX instruction set), and $32\,\mathrm{Kb}$ of L1 and $256\,\mathrm{Kb}$ of L2 caches. In the introductory example examined thoroughly in Sec.\,\ref{Sec:MaximisingPerformanceCPU}, by increasing the unroll factor, the physical register file filled-up first, then the application started to extensively use the L1 cache as well. This can be clearly seen in Tab.\,\ref{ProfIntroductoryExample} in the increase of the L1 cache loads by increasing the unroll factor from $1$ (second column) to $8$ (first column). As long as the whole computational problem fits into the L1 and/or L2 caches, the performance is usually not limited by memory bandwidth (this the case for all the test problems examined in the present paper).

In contrast to CPUs, GPUs follow the memory throughput-oriented design. In one hand, it uses a huge register file per streaming multiprocessor: $65536$ entries of $32$ bit registers (Kepler or newer architectures), which is considered significant even though a streaming multiprocessor handles a massive number of threads (e.g., a maximum of $2048$ in Kepler architecture). An equivalent number for the Ivy Bridge CPU architecture (single core) is $144 \cdot 8=1152$ entries of $32$ bit registers. On the other hand, a GPU has a small amount of L1 cache or shared memory (user-programmable L1 cache); their total size for hundreds or thousands of threads is in the order of the L1 and L2 caches of the CPU (for a maximum of $2$ threads). In the case of the Kepler architecture, their total size is $64\,\mathrm{Kb}$ (GK110) or $128\,\mathrm{Kb}$ (GK210). In proportion, the amount of its last-level L2 cache is even smaller: $1536\,\mathrm{Kb}$ for tens or hundreds of thousands of threads (the unified last-level L3 cache size of Ivy Bride is $2-8\,\mathrm{Mb}$ for $2-8$ threads). Thus, the strategy of the GPU to hide latency is to use a massive number of threads with a large register file. The context switch between the warps has no delay; consequently, the streaming multiprocessor has a high flexibility to select some eligible warps for execution, while some others are waiting for data arriving from the global memory. Because of the large number of registers, a single warp can usually perform many instructions before stalling due to a data request. As the decrease of the latency is not a high priority, the throughput of the global memory bandwidth can be increased drastically. In case of a simple low order system, like the Lorenz equation or the introductory example, the complete problem fits into the registers, and the application becomes extremely fast and free of global memory transactions.

Although the amount of registers per thread is relatively high, it is still a scarce resource and the major limiting factor for the maximum number of residing threads in an SM. Keep in mind that more residing threads mean more residing warps to hide latency. In one extreme, using the maximum allowed $255$ registers per threads, the maximum number of residing threads in an SM is $65536/255 \approx 256$. At the other extreme, for $2048$ residing threads (hardware limit of Kepler), the number of registers per threads should not be higher than $65536/2048=32$. The maximum number of registers per thread can be managed by the compiler option
\begin{verbatim}
--maxrregcount.
\end{verbatim}
Another important (hardware) limiting factor is the maximum number of residing blocks in an SM ($16$ in Kepler). By setting the block size to $32$ threads (warp size), the maximum number of residing threads is $16 \cdot 32=512$. The ratio of the theoretical maximum of the residing threads and the maximum allowed by the hardware limitation is called occupancy $O$, e.g., $O=512/2048=0.25=25\%$. There are many other resource-related or hardware limiting factors that affect the achievable occupancy. Its calculation is highly non-trivial; thus, the user is referred to the official occupancy calculator of Nvidia \cite{Occupancy_Calculator}.

In general, for latency-limited applications, the basic strategy is to increase the occupancy. Higher number of residing warps means a higher possibility of hiding latency. However, when the block size is already large enough, the occupancy can only be increased by decreasing the maximum registers per threads. If the available registers per thread are insufficient, some variables are ``spilt'' back to global memory further increasing pressure on global memory. Finding an optimal setup (maximum registers, block size and the total number of threads) is not trivial and highly problem dependent. For the test cases presented in the forthcoming sections, the compiler options and other details are provided. In ODEINT and DifferentialEquations.jl, most of these cannot be managed by the user, while in MPGOS they can be tuned explicitly.

In some ODE systems (other than the test cases presented here), the on-chip shared memory of the GPU (user-programmable cache) can also be exploited. Moreover, many other issues can affect the final performance of the application (e.g., frequent PCI-E memory transactions), for the details, see our previous publication \cite{Hegedus2020b} and the manual of MPGOS \cite{Hegedus2019a}.

\subsubsection{Maximising FLOPS Efficiency} \label{Sec:MaximisingFLOPSEfficiency}

In this section, the effect of the total number of threads and the block size of a GPU simulation is demonstrated on the introductory example defined by Eq.\,\eqref{exemplary_system}. The structure of the kernel function is very similar to the one shown in Lst.\,\ref{NaiveApproach}; thus, it is not repeated here. The full code can be found in the corresponding GitHub repository \cite{Basic_tests_RK4_GitHub}. No special care is taken for unrolling and explicit vectorisation in the program level: it follows the na\"ive approach. The latency is handled by the execution configuration (block size) and the register usage. This kernel function is called by every thread and executed independently but with different parameter values (selected by a unique thread identifier). Compiling the code with the option
\begin{verbatim}
--ptxas-options=-v,
\end{verbatim}
the register usage per thread is printed, which is only $16$ in this case ($32$ is enough for $O=100\%$ occupancy). Therefore, it is a perfect example of how to deal with latency via the block size and the total number of threads when the number of registers is not an issue. 

First of all, to distribute the workload to the streaming multiprocessors (SM) evenly, the block size $B$ and the total number of threads $N$ (also the total number of instances of the ODE system solved simultaneously) have to be properly selected. The introductory example is tested on an Nvidia Titan Black GPU that have $N_{SM}=15$ SMs. To distribute the same amount of blocks to each SM, the total number of threads have to be $N=B \cdot N_{SM} \cdot i$, where $i$ is an arbitrary positive integer; otherwise, some SMs will be idle near the end of the simulations. The block size $B$ should be an integer multiple of the warp size $32$; otherwise, during the execution of the last warp, some compute units will be idle. Five configurations are tested, and their profiling data is summarized in Tab.\,\ref{ProfIntroductoryExampleGPU}.

\begin{table}
  \caption{Summary of the condensed representation of the profiling information for the introductory example. The peak double-precision processing power of the employed Nvidia GeForce Titan Black GPU is 1707 GFLOPS.}
  \label{ProfIntroductoryExampleGPU}
  \begin{equation*}
    \begin{array}{|c|ccccc|}
      \hline
     \text{B (block size)} & \text{32} & \text{32} & \text{32} & \text{64} & \text{128} \\
     \text{N (total threads)} & \text{480} & \text{7680} & \text{61440} & \text{61440} & \text{61440} \\
     \hline \hline
     \text{Runtime [$\mu$s]} & 114.56 & 125.67 & 961.42 & 918.00 & 868.52 \\
     \text{SM activity [\%]} & 95.03 & 94.83 & 98.53 & 94.48 & 97.14 \\
     \text{Achieved occup.} & 0.0156 & 0.245 & 0.249 & 0.460 & 0.878 \\
     \text{Elig. warps} & 0.24 & 4.8 & 4.9 & 17.0 & 42.1 \\
     \text{Efficiency [$\%$]} & 4.97 & 74.79 & 79.25 & 84.62 & 86.86 \\
     \text{Exec. dep. [\%]} & 91.03 & 83.13 & 83.11 & 47.84 & 24.92 \\
     \text{Not sel. [\%]} & 2.52 & 10.25 & 10.41 & 48.31 & 71.02 \\ \hline
    \end{array}
  \end{equation*}
\end{table}

To acquire the profiling data, the Nvidia profiler tool \textbf{nvprof} is employed. It enables one to collect events and metrics for kernel functions from the command line. A typical output for a few metrics are presented in Lst.\,\ref{OutputNvprof}. The discussion of the full capabilities of \textbf{nvprof} is beyond the scope of the present paper, the employed shell script and the complete profiling data can be found in the GitHub repository \cite{Basic_tests_RK4_GitHub} of the problem. With the profiler, one can obtain several useful pieces of information directly. For instance, the distribution of floating-point addition, multiplication and fused multiply-add instructions. From them, the maximum possible floating-point efficiency (exploitation of the peak performance in percentage) can be calculated: $95\%$. In addition, the achieved floating-point efficiency can be measured directly (82.36\%). The utilisation of the memory subsystem (cache hierarchy) and the different compute units are important metrics to determine what is the main performance-limiting factor of the application (memory bandwidth, compute units or latency). Next, the stall reasons can further narrow the focus to solve the most severe bottleneck. Finally, the multiprocessor activity is a good measure of how evenly the blocks are distributed to the SMs.

\begin{lstlisting}[float, basicstyle=\small\selectfont\ttfamily, label=OutputNvprof, caption=Typical output of a profiling with the utility \textbf{nvprof}.]
Multiprocessor Activity                              96.41%
Achieved Occupancy                                 0.402929
Eligible Warps Per Active Cycle                   13.914399
Floating Point Operations(Double Precision Add)    15360000
Floating Point Operations(Double Precision Mul)           0
Floating Point Operations(Double Precision FMA)   153600000
Integer Instructions                                3932160
FLOP Efficiency(Peak Double)                         82.36%
L1/Shared Memory Utilization                        Low (1)
L2 Cache Utilization                                Low (1)
Device Memory Utilization                           Low (1)
Load/Store Function Unit Utilization                Low (1)
Arithmetic Function Unit Utilization               Max (10)
Issue Stall Reasons (Pipe Busy)                       0.66%
Issue Stall Reasons (Execution Dependency)           52.36%
Issue Stall Reasons (Data Request)                    0.26%
Issue Stall Reasons (Not Selected)                   43.02%
\end{lstlisting}

Table\,\ref{ProfIntroductoryExampleGPU} summarises some key metrics for the five different kernel launch configurations of the introductory example given by Eq.\,\eqref{exemplary_system}. As a reminder, the task is to perform $1000$ steps with the classic fourth-order Runge--Kutta scheme on an ensemble of $N$ ODE systems. Applying a block size of $B=32$, the theoretical occupancy is only $0.25=25\%$. Observe that the runtimes in the first two columns have minor difference even though the total number of solved systems differ by more than an order of magnitude. This is a perfect example that enough threads have to be launched to fully utilise the GPU. The runtime scales linearly approximately above ($N=7680$), compare columns $2$ and $3$. That is, the utilisation of the GPU is saturated, which can also be seen from the achieved occupancy. It is close to the theoretical one. Although the floating-point efficiency is already high (above $70\%$), the stall reasons due to execution dependency still above $80\%$ indicating latency-limited behaviour caused by the arithmetic instructions. The ``remedy'' is to increase the occupancy and hide latency. It must be stressed that a high percentage of a stall reason does not necessarily indicate poor performance as the number of the stalled cycles can still be small. It is demonstrated by the high floating-point efficiencies in the second and third columns in Tab.\,\ref{ProfIntroductoryExampleGPU}. The major limiting factor of the occupancy is the maximum number of blocks per SM ($16$): $O=16*32/2048=0.25=25\%$. In the last two columns of Tab.\,\ref{ProfIntroductoryExampleGPU}, the block size is increased to $64$ and $128$ to increase occupancy to $50\%$ and $100\%$, respectively. The latency due to execution dependency gradually disappears, and the major stall reason becomes the ``Issue Stall Reasons (Not Selected)''. This stall reason means that there are more eligible warps in the SM to perform instruction execution than compute units, and the warp scheduler chooses a different warp to execute. Thus, high floating-point efficiency and high stall reason of ``Not Selected'' implies high performance.

\subsection{The effect of mathematical functions and division} \label{Sec:SpecialFunctionsDivisions}

In order to ``feel'' the consequences of the usage of division and special functions, two additional micro-benchmark examples are provided. They are tested only on a CPU (Ivy Bridge); however, the conclusions are valid for GPUs as well. These operations are extremely expensive compared to addition/subtraction and multiplication. To test the effect of division, Eq.\,\eqref{exemplary_system} is modified to
\begin{equation} \label{exemplary_system_division}
\dot{x} = \frac{1}{x} - p \cdot x.
\end{equation}
The corresponding profiling results are shown in the last column of Tab.\,\ref{ProfIntroductoryExample}. The divider is almost $100\%$ active, the floating-point efficiency dropped down to $11\%$ and the runtime is almost an order of magnitude higher than the optimal code (first column) for Eq.\,\eqref{exemplary_system}. The optimal unroll factor here is $2$, and the VCL library is still used for explicit vectorisation. The cache loads are relatively low; thus, latency by memory is not an issue. On the other hand, the stall due to execution dependency is very high (last row). The reason is the high latency of the division instruction (approximately $21-45$ clock cycles). In addition, the divider blocks port-0 for performing multiplication. \textit{The conclusions also hold for general integer division.} For a detailed discussion, see again Sec.\ref{Sec:CPUArchitectureProfiling}.

To test the effect of mathematical functions, Eq.\,\eqref{exemplary_system} is modified again as follows
\begin{equation} \label{exemplary_system_transcendental}
\dot{x} = p \sin(x)
\end{equation}
to include the transcendental function sine. The column with the header \textit{transc. vcl unroll} depicts the profiling results. Although the runtime is more than two times higher than in the case of division, the floating-point efficiency is much higher ($37.3\%$). The reason is the way the hardware handles special functions. Even if the instruction set supports the evaluation of such a function, the computation process is decomposed into several micro-operations. For instance, Intel uses polynomial based approximations to compute trigonometric functions. Therefore, the higher runtime is due to the increased number of required instructions, see again Tab.\,\ref{ProfIntroductoryExample}. In case of no hardware support of a special function, external libraries have to be used to approximate by software. Another important issue is the exploitation of vector registers. Fortunately, the VCL library supports the evaluation of mathematical functions (transcendental, power or root). It is claimed that the vectorised versions are much faster than most of the commonly used scalar alternatives \cite{Fog2020b}. The effect of the increased number of operations and the larger number of required intermediate variables can also be seen in the very high number of L1 cache loads. This test has the highest pressure on the L1 cache. Thus, the stall reasons are mainly due to the L1 cache latency.

Naturally, in many cases, the use of division or mathematical functions are necessary. However, one can still try to minimise their usage. For instance, during the right-hand side evaluation of the Keller--Miksis equation (Sec.\,\ref{Sec:KellerMiksisEquation}), the reciprocal of the variable $x_1$ is computed in advance and used as a multiplicator in several places. In the same example, the calculation of both the sine and cosine values of a variable can be done together efficiently. Many programming languages and libraries provide functions that compute both the sine and cosine values approximately within the same time as a single trigonometric function evaluation. Moreover, if it is possible, replace a general power function having a rational exponent to a combination of square and reciprocal square functions and multiplications/divisions. It must be stressed that the issues introduced here are not specific to the employed numerical algorithm. \textit{The efficient implementation of the right-hand side of the ODE is under the control of the user alone.} For more detailed instructions and guidance, the reader is referred to the documentation of the VCL library \cite{Fog2020b} for CPUs, and to the CUDA documentation \cite{CUDA_BestPracticeGiude_InstructionOptimisation} in case of GPUs.

\section{Performance characteristics of the test cases} \label{Sec:PerformanceCharacteristics}

The introductory example in Sec.\,\ref{Sec:IntroductoryExample} was ideal for demonstrating possible performance issues, and investigating basic optimisation techniques. In this section, the performance characteristics of more complex non-linear models are presented, where the runtimes are plotted as a function of the number of the instances of the solved ODE systems. Detailed profiling is omitted here, and the runtimes provide the bases for the comparison of the different program packages. Nevertheless, whenever it is feasible, the techniques revealed by the detailed profiling in Sec.\,\ref{Sec:IntroductoryExample} are employed to maximise performance. For CPU solvers (ODEINT, DifferentialEquations.jl), this means explicit vectorisation and instruction-level parallelism (ILP) by unroll. Even though all the studied program packages are able to shuttle computations to GPUs, only in the case of MPGOS has the user the option to tune the register usage and the size of the thread blocks. In the case of DifferentialEquations.jl, they are completely hidden by the underlying abstraction. In our ODEINT versions, the ODE functions are implemented in a CUDA file using the Thrust library that must be compiled with \textbf{nvcc}. Thus, at least the maximum register usage can be managed. Similarly to the Standard Template Library (STL) in C++, Thrust provides many containers and algorithms that can be run on CUDA supported GPUs. Thrust is part of the NVIDIA CUDA framework \cite{CUDA_Thrust}. In the subsequent sections, three different models are examined with large alteration in complexity and features that need to be handled. During the simulations, only double-precision floating-point arithmetics is used.

All of our source codes can be found in a GitHub repository \cite{ODE_Solver_Tests_GitHub}. The applications are tested under Ubuntu 20 LTS operating system. The C++ compiler is gcc 7.5.0, and the CUDA Toolkit version is 10.0. The versions of the program packages is as follows; ODEINT: v2, MPGOS: v3.1, Julia: v1.5.0 and DifferentialEquations.jl: v6.15.0. In case of Julia, many other modules are used: DiffEqGPU v1.6.0, SimpleDiffEq v1.2.1, LoopVectorization v0.8.24.

\subsection{The Lorenz system} \label{Sec:LorenzSystem}

The first test model is the classic Lorenz system written as
\begin{align} \label{Lorenz1}
\dot{x}_1  &= 10.0 \, (x_2-x_1),\\ \label{Lorenz2}
\dot{x}_2  &= p \, x_1 - x_2 - x_1 \, x_3, \\ \label{Lorenz3}
\dot{x}_3  &= x_1 \, x_2 - 2.666 \, x_3,
\end{align}
composed of three first-order equations, where $x_i$ are the components of the state space, $p$ is a parameter, and the dot stands for the time derivative. The system was first studied by Edward Lorenz for simplified modelling of the atmospheric convection \cite{Lorenz1963a}. The system is famous for its chaotic solution, the Lorenz attractor. It has been the subject of hundreds of research articles \cite{Immler2018a,Graca2018a,Meucci2008a,Goswami2008a,Goswami2007a,Stewart2000a}; also, a complete book is devoted to the study of the Lorenz equations and their non-linear dynamics \cite{Sparrow1982a}.

During the computations, the control parameter $p$ is varied between $0.0$ and $21.0$ with a resolution $N$ distributed uniformly. That is, during a single run, $N$ instances of Eqs.\,\eqref{Lorenz1}-\eqref{Lorenz3} are solved each having a different parameter value. The main objective is to calculate the runtime of $1000$ time steps with the classic $4^{th}$ order Runge--Kutta method (fixed time steps) as a function of $N$. These performance characteristic curves obtained by using the aforementioned program packages (using both CPUs and GPUs) are compared and examined.

Due to its simplicity, the Lorenz system is a standard test case for many program packages. Note that the right-hand side involves only additions/subtractions and multiplications. In addition, because of the fixed time steps, there is no thread divergence (every instance needs exactly the same number of steps). Therefore, the total amount of work is well-defined, which makes this system a perfect example to test the vectorisation capabilities of the program codes (AVX in CPUs) and the exploitation of the raw peak processing power of the hardware.

\subsubsection{Performance curves on CPUs} \label{Sec:LorenzPerformanceCurvesCPU}

\begin{figure}[ht!]    
	\centering
		\includegraphics[width=17.0cm]{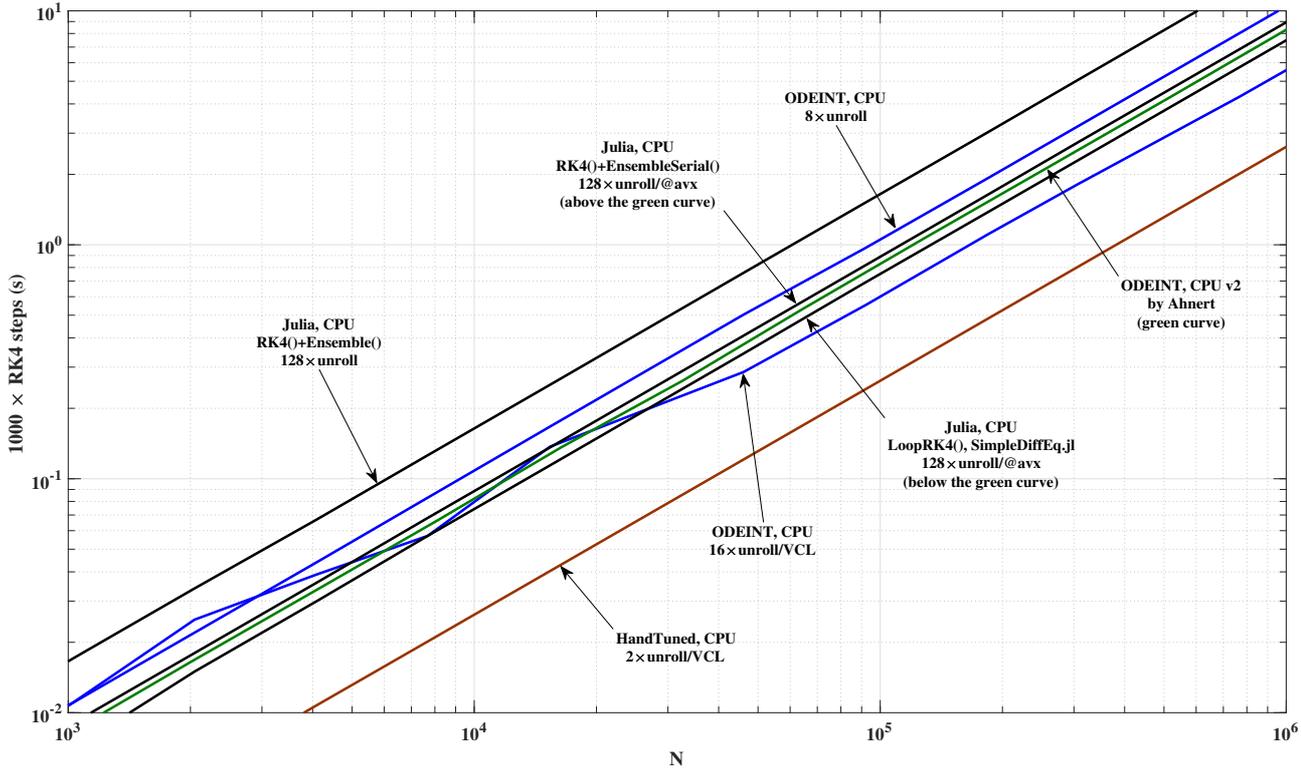}
	\caption{Performance curves of the Lorenz system; that is, the runtime of $1000$ steps with the classic $4^{th}$ order Runge--Kutta method as a function of the ensemble size $N$. Brown curve: our ``hand-tuned'' version written in C++; blue curves: ODEINT; green curve: ODEINT by Ahnert et al., \cite{Ahnert2014a}; black curves: Julia (DifferentialEquations.jl).}
	\label{Fig:LorenzPerformanceCurvesCPU}
\end{figure}

The performance curves obtained on CPUs are summarised in Fig.\,\ref{Fig:LorenzPerformanceCurvesCPU}. The blue and black curves are computations with ODEINT and DifferentialEquations.jl (Julia for short), respectively. The brown curve is our ``hand-tuned'' version written in C++. Finally, the green performance characteristic curve is digitalised from the publication written by the developers of ODEINT \cite{Ahnert2014a} (Karsten Ahnert and his co-workers). The simulations (except the green ones) are performed on an Intel Core i7-4820K CPU using a single core having a $30.4$ GFLOPS peak double-precision performance. For the computations by Ahnert et al., an Intel Core i7-920 CPU is employed using all of the $4$ cores; the total peak double-precision performance is $42.56$ GFLOPS. Code snippets are omitted during the discussion as all the source codes can be found in the GitHub repository \cite{Lorenz_RK4_GitHub}.

The fastest solver is our own ``hand-tuned'' implementation in C++ (brown curve), which follows the instruction-level parallelism (ILP) described by Lst.\,\ref{InstructionLevelParallelism} in Sec.\,\ref{Sec:InstructionLevelParallelism} (the optimal unroll factor is $2$), and it exploits the explicit vectorisation possibility via the VCL library, see Sec.\,\ref{Sec:ExplicitVectorisation}. This code is specialised only for the Lorenz system; thus, it is free of any overhead that stems from constraints of general-purpose solvers. This case serves as a reference for comparison with other program packages, and it has approximately a $70\%$ floating-point efficiency.

ODEINT provides the fastest general-purpose alternatives (blue and green curves). Employing only the unroll technique to increase ILP (optimal unroll factor is $8$, upper blue curve), the code is approximately $4$ times slower than the reference computation. It is very likely that the compiler cannot generate a code to exploit the vectorisation capabilities of the CPU automatically. Therefore, the results of the brown and the upper blue curves are consistent as with full vectorisation, the maximum speed-up is exactly a factor of $4$. Interestingly, for the ODEINT implementation, a larger unroll factor was necessary to hide the latency as much as possible. In the case of ODEINT, the only possibility to increase ILP is to pack multiple Lorenz systems into a single ODE function in a similar way as demonstrated by Eq.\,\eqref{array_exemplary_system} (the internals of the solver are hidden from the user). The results taken from \cite{Ahnert2014a} indicate a somewhat faster implementation (green curve). However, the total processing power of the used CPU is $42.56$ GFLOPS instead of $30.4$ GFLOPS. Assuming a linear correlation between the runtimes and the peak processing power and employing a correction factor of $42.56/30.4=1.4$ on the green curve, the runtimes of the upper blue and the corrected green curves are within $10\%$ difference. These results also imply consistency. One of the main advantages of ODEINT is its compatibility with different data structures. Therefore, the explicit vectorisation by the VCL library can easily be implemented by replacing the state type from double to Vec4d. In this case, the optimal unroll factor is increased to $16$, see the lower blue curve. Including the explicit vectorisation, the speed-up is only $\times 1.75$ between the two blue curves in the asymptotic regime ($N>40000$), which indicates a suboptimal usage of the vector registers. In summary, ODEINT is a fast solver with CPUs; its only drawback is that the vectorisation capabilities cannot be fully exploited.

In general, Julia provides a somewhat slower option to the Lorenz problem. By employing the unroll technique alone, the code is approximately $6.3$ times slower (uppermost black cure) than the baseline ``hand-tuned'' C++ version (brown curve). This is larger than the optimal $4$ times slow down (``hand-tuned'' without vectorisation). This implies the presence of some amount of overhead compared to ODEINT, which is also confirmed by the much larger required unroll factor ($128$) to hide latency. Similarly to ODEINT, the ILP can only be increased by packing multiple Lorenz systems into a single ODE function. The vectorisation can be employed during the evaluation of the right-hand side simply by using the @avx macro (LoopVectorization.jl package) on the loop inside the ODE function. The speed-up is $\times 1.85$ (middle black curve) that is lower than the optimal ($\times 4$), but is still larger than the $\times 1.75$ speed-up produced by ODEINT. The runtime can further be decreased by using the LoopRK4() solver from the package SimpleDiffEq.jl (lowermost black curve). This version of the code is still $2.9$ times slower than the baseline, and $1.3$ times slower than the fastest ODEINT code (lower blue curve). \textit{Although Julia is somewhat slower than ODEINT, it is a high-level programming language in which the code development is extremely easy and fast compared to ODEINT.}

\subsubsection{Performance curves on GPUs} \label{Sec:LorenzPerformanceCurvesGPU}

\begin{figure}[ht!]    
	\centering
		\includegraphics[width=17.0cm]{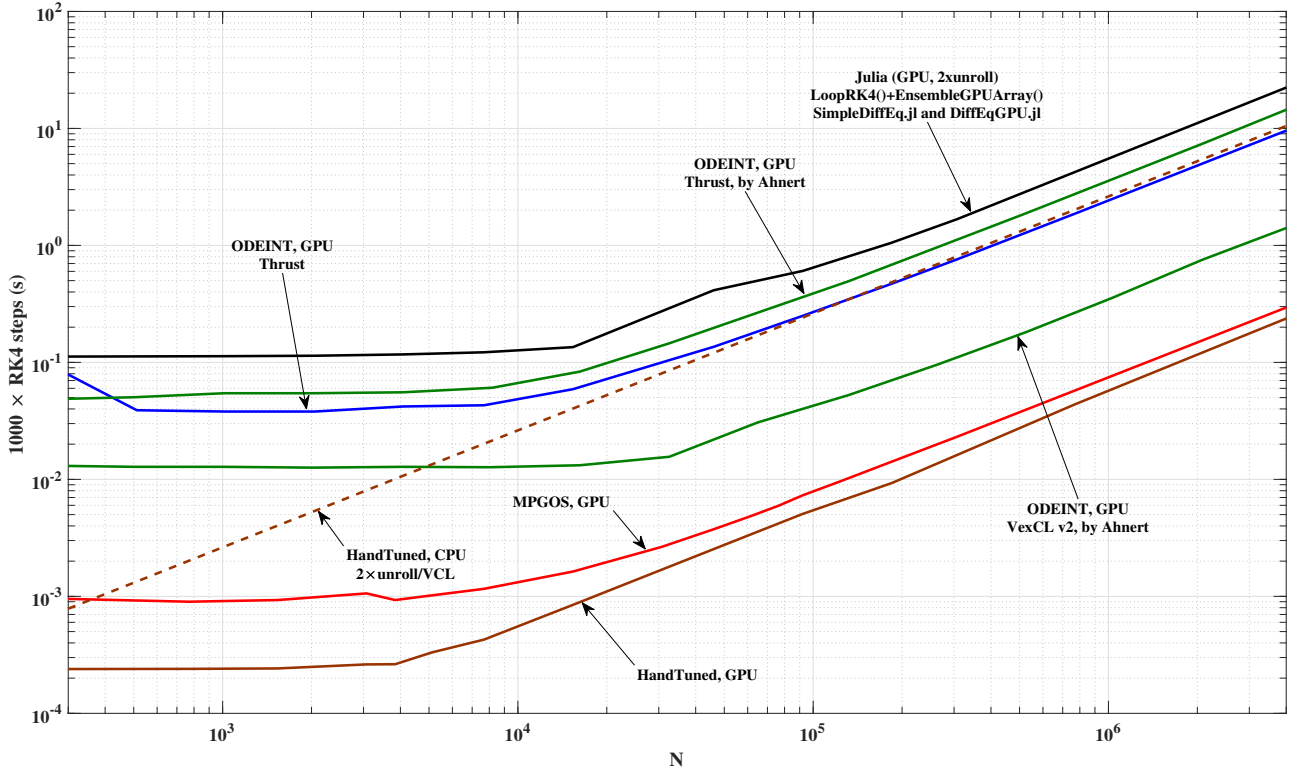}
	\caption{Performance curves of the Lorenz system; that is, the runtime of $1000$ steps with the classic $4^{th}$ order Runge--Kutta method as a function of the ensemble size $N$. Brown curves: our ``hand-tuned'' versions written in C++ or CUDA C; blue curve: ODEINT; green curves: ODEINT by Ahnert et al., \cite{Ahnert2014a}; black curve: Julia (DifferentialEquations.jl).}
	\label{Fig:LorenzPerformanceCurvesGPU}
\end{figure}

The performance curves obtained on GPUs are summarised in Fig.\,\ref{Fig:LorenzPerformanceCurvesGPU} together with the ``hand-tuned'' baseline CPU solver. The colour code is the same as in the case of Fig.\,\ref{Fig:LorenzPerformanceCurvesCPU}. Namely, the blue and black curves are computations with ODEINT and Julia, respectively. The brown curves are our ``hand-tuned'' versions written in C++ (CPU, dashed) and CUDA C (GPU, solid). Finally, the green performance characteristic curves are again digitalised from the publication written by the developers of ODEINT \cite{Ahnert2014a}. The simulations (except the green ones) are performed on an Nvidia Titan Black GPU having a $1707$ GFLOPS peak double-precision performance. During the computations by Ahnert et al., an Nvidia Tesla K20c GPU was employed, which has a peak double-precision performance of $1175$ GFLOPS. Code snippets are omitted during the discussion as all the source codes can be found in the GitHub repository \cite{Lorenz_RK4_GitHub}.

The fastest baseline simulation is shown by the solid brown curve that is free of any overhead. Profiling with \textbf{nvprof}, both the optimal block size and maximum register usage (compiler option) are $64$. The actual register usage is $40$; thus, the $50\%$ theoretical occupancy can easily be achieved. The floating-point efficiency and also the ``Issue Stall Reasons (Not Selected)'' are above $80\%$ in the asymptotic regime (approximately above $N=40000$). Therefore, the code is considered to be highly optimized. The folder ``ProfileData'' of the related GitHub repository contains detailed profiling results for $N=1536$, $15360$ and $768000$. This solution is $50$ times faster than the baseline CPU solver (dashed brown curve). The GPU, however, has $56$ times higher double-precision peak performance than the CPU. This small discrepancy can come from the technique the floating-point efficiency is obtained for CPUs. It must be calculated indirectly from the number of arithmetic instructions and the runtime. In addition, the very low runtime of the GPU computations (e.g., $40\,\mathrm{ms}$ at $N=768000$) can lead to high relative uncertainties, as a small change in the runtime can cause a large speed-up difference.

The performance characteristic curve of the program package MPGOS is close to the baseline simulations in the asymptotic regime where the arithmetic units of the GPU are saturated with enough instructions, and the GPU is fully utilised. The slow down is approximately $1.3$ times. This $30\%$ increase in runtime is the price of being general purpose and modular. For low values of $N$, the difference is much larger (close to a $4$ times slow down). However, this range of $N$ represents suboptimal, underutilised usage of the GPUs. Therefore, it has less practical relevance.

The Julia GPU solver (black curve) and the ODEINT implementations with the Thrust library (blue and upper green curves) have similar performance characteristics. All have poor performance. They have higher or have nearly equal runtimes compared to our baseline CPU solver, although the peak performance of the underlying hardware is more than an order of magnitude higher. The common approach of these implementations is the separate invocation of the kernel function(s) at each time step. That is, at the beginning of every time step, all the variables have to be loaded from the slow global memory of the GPU to the registers again. Therefore, there is no chance for data reuse and latency hiding via the fastest register memory, and the applications are memory bandwidth limited. Keep in mind that the latency of the global memory is as high as approximately $600$ clock cycles.

Ahnert and his co-workers in their publication \cite{Ahnert2014a} provide an alternative solution that builds-up a single monolithic kernel using the library VexCL \cite{VexCL_GitHub}. The VexCL library is similar to Thrust, but it supports a variety of hardware (Nvidia and AMD/ATI GPUs, multi-core CPUs). The corresponding digitalised runtimes are shown by the lower green curve in Fig.\,\ref{Fig:LorenzPerformanceCurvesGPU}. The performance of this case is significantly improved by eliminating the memory bandwidth limiting bottleneck. The runtime difference between this ODEINT version and MPGOS is reduced to $\times 4.6$. However, according to the authors, this ODEINT approach has severe restrictions: \textit{``it only supports embarrassingly parallel problems (no data dependencies between threads of execution), and it does not allow conditional statements or loops with non-constant number of iterations''}. Therefore, the flexibility of the code has been lost, which is important in many situations, see Sec.\,\ref{Sec:KellerMiksisEquation} or Sec.\,\ref{Sec:ImpactDynamics}.

\subsection{The Keller--Miksis equation} \label{Sec:KellerMiksisEquation}

The second test model is the Keller--Miksis equation describing the evolution of the radius of a spherical gas bubble placed in a liquid domain and subjected to external excitation \cite{Lauterborn2010a}. It is a non-linear second-order ordinary differential equation. In order to remain consistent with our previous publications \cite{Hegedus2020a,Hegedus2018a,Hegedus2020c}, the dual-frequency version is employed. The system reads as
\begin{equation}\label{keller_miksis_1}
\left( 1-\frac{\dot{R}}{c_L} \right) R\ddot{R} + \left( 1-\frac{\dot{R}}{3c_L} \right) \frac{3}{2} \dot{R}^2 = \left( 1+\frac{\dot{R}}{c_L} + \frac{R}{c_L}\frac{d}{dt} \right) \frac{\left( p_L - p_{\infty}(t) \right)}{\rho_L},
\end{equation}
where $R(t)$ is the time dependent bubble radius; $c_L=1497.3\,\mathrm{m/s}$ and $\rho_L=997.1\,\mathrm{kg/m^3}$ are the sound speed and density of the liquid domain, respectively. The dot stands for the derivative with respect to time. The pressure far away from the bubble, $p_{\infty}(t)$, is composed of static and periodic components
\begin{equation}\label{keller_miksis_2}
p_{\infty}(t) = P_{\infty} + P_{A1} \sin(\omega_1 t) + P_{A2} \sin(\omega_2 t + \theta),
\end{equation}
where $P_{\infty}=1\,\mathrm{bar}$ is the ambient pressure. The periodic components have pressure amplitudes $P_{A1}$ and $P_{A2}$, angular frequencies $\omega_1=2\pi f_1$ and $\omega_2=2\pi f_2$, and a phase shift $\theta$.

The connection between the pressures inside and outside the bubble at its interface is chosen as
\begin{equation}\label{keller_miksis_3}
p_G + p_V = p_L + \frac{2 \sigma}{R} + 4 \mu_L \frac{\dot{R}}{R},
\end{equation}
where the total pressure inside the bubble is the sum of the partial pressures of the non-condensable gas, $p_G$, and the vapour, $p_V=3166.8\,\mathrm{Pa}$. The surface tension is $\sigma=0.072\,\mathrm{N/m}$ and the liquid kinematic viscosity is $\mu_L=8.902^{-4}\,\mathrm{Pa\,s}$. The gas inside the bubble is assumed to obey a simple polytropic relationship
\begin{equation}\label{keller_miksis_4}
p_G = \left( P_{\infty} - p_V + \frac{2 \sigma}{R_E} \right) \left(\frac{R_E}{R}\right)^{3 \gamma},
\end{equation}
where the polytropic exponent $\gamma$ for air is chosen ($\gamma=1.4$, adiabatic behaviour) and the equilibrium bubble radius is $R_E$.

System\,\eqref{keller_miksis_1}-\eqref{keller_miksis_4} is transformed into a dimensionless form by the introduction of the following dimensionless variables
\begin{align}\label{dimensionless_time}
\tau &= \frac{\omega_1}{2 \pi} t,\\
y_1  &= \frac{R}{R_E},\\
y_2  &= \dot{R} \frac{2\pi}{R_E \omega_1}.
\end{align}
The dimensionless system is written as
\begin{align}\label{dimensionless_system1}
\dot{y}_1  &= y_2,\\\label{dimensionless_system2}
\dot{y}_2  &= \frac{N_{\mathrm{KM}}}{D_{\mathrm{KM}}},
\end{align}
where the numerator, $N_{\mathrm{KM}}$, and the denominator, $D_{\mathrm{KM}}$, are
\begin{multline}\label{numerator}
N_{\mathrm{KM}} = \left( C_0 + C_1 y_2 \right) \left( \frac{1}{y_1} \right)^{C_{10}} - C_2 \left( 1 + C_9 y_2 \right) - C_3 \frac{1}{y_1} - C_4 \frac{y_2}{y_1} - \\
\left( 1 - C_9 \frac{y_2}{3} \right) \frac{3}{2} y_2^2 
-\left( C_5 \sin(2 \pi \tau) + C_6 \sin(2 \pi C_{11} \tau + C_{12}) \right) \left( 1 + C_9 y_2 \right) \\
-y_1 \left( C_7 \cos(2 \pi \tau) + C_8 \cos(2 \pi C_{11} \tau + C_{12}) \right),
\end{multline}
and
\begin{equation}\label{denominator}
D_{\mathrm{KM}} = y_1 - C_9 y_1 y_2 + C_4 C_9,
\end{equation}
respectively.

During the rearrangement of the dimensionless equations, special care was taken to minimise the computational effort during the evaluation of the right-hand side. Therefore, the following system coefficients, which can be computed in advance of the integration, are extracted:
\begin{align}\label{coefficients}
C_0 &= \frac{1}{\rho_L} \left( P_{\infty} - p_V + \frac{2 \sigma}{R_E} \right) \left( \frac{2 \pi}{R_E \omega_1} \right)^2,\\
C_1 &= \frac{1-3\gamma}{\rho_L c_L} \left( P_{\infty} - p_V + \frac{2 \sigma}{R_E} \right) \frac{2 \pi}{R_E \omega_1},\\
C_2 &= \frac{P_{\infty} - p_V}{\rho_L} \left( \frac{2 \pi}{R_E \omega_1} \right)^2,\\
C_3 &= \frac{2 \sigma}{\rho_L R_E} \left( \frac{2 \pi}{R_E \omega_1} \right)^2,\\
C_4 &= \frac{4 \mu_L}{\rho_L R_E^2} \frac{2 \pi}{\omega_1},\\
C_5 &= \frac{P_{A1}}{\rho_L} \left( \frac{2 \pi}{R_E \omega_1} \right)^2,\\
C_6 &= \frac{P_{A2}}{\rho_L} \left( \frac{2 \pi}{R_E \omega_1} \right)^2,\\
C_7 &= R_E \frac{\omega_1 P_{A1}}{\rho_L c_L} \left( \frac{2 \pi}{R_E \omega_1} \right)^2,\\
C_8 &= R_E \frac{\omega_1 P_{A2}}{\rho_L c_L} \left( \frac{2 \pi}{R_E \omega_1} \right)^2,\\
C_9 &= \frac{R_E \omega_1}{2 \pi c_L},\\
C_{10} &= 3\gamma,\\
C_{11} &= \frac{\omega_2}{\omega_1},\\
C_{12} &= \theta.
\end{align}

Assuming fixed liquid ambient properties (temperature $T_{\infty}=25\,\mathrm{C^0}$ and pressure $P_{\infty}=1\,\mathrm{bar}$) and polytropic exponent, six control parameters remain in the system. The properties of the external forcing: pressure amplitudes $P_{A1}$ and $P_{A2}$, frequencies $\omega_1=2 \pi f_1$ and $\omega_2=2 \pi f_2$, and the phase shift between the harmonic components of the driving $\theta$. Also, the bubble size (equilibrium radius $R_E$) is the sixth parameter. Note, however, that from the implementation point of view, the number of the parameters of the system is $13$ ($C_0$ to $C_{12}$). Although the usage of the coefficients $C_0-C_{12}$---instead of the six ``real'' parameters---requires additional storage capacity and global memory load operations, it can significantly reduce the number of the necessary computations as these coefficients are pre-computed (always on the CPU).

\begin{figure}[ht!]    
	\centering
		\includegraphics[width=8.6cm]{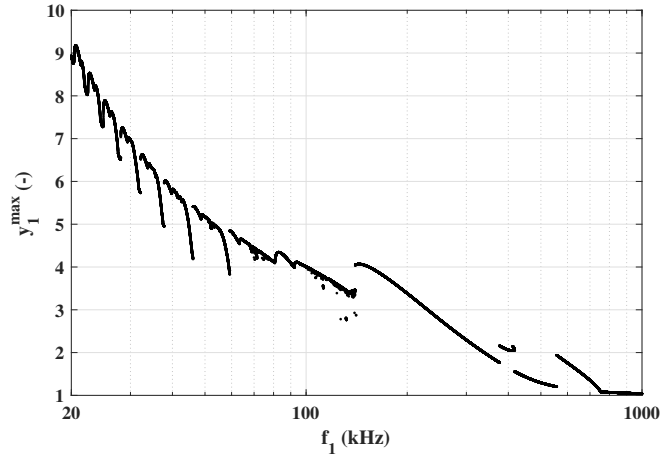}
	\caption{Frequency response diagram of the Keller--Miksis equation where the maximum value of the dimensionless bubble radius $y_1^{max}$ is plotted as a function of the frequency $f_1$. The value of $f_1$ is varied between $20\,\mathrm{kHz}$ and $1\,\mathrm{MHz}$ with a resolution $N$ distributed logarithmically.}
	\label{Fig:FrequencyResponseCurve}
\end{figure}

The main aim of this study is examining the numerical behaviour of the Keller--Miksis equation instead of performing detailed parameter studies with exhaustive physical interpretation. Thus, only the first frequency component $f_1$ of the external driving is chosen as a control parameter and the rest is kept fixed ($P_{A1}=1.5\,\mathrm{bar}$, $P_{A2}=0\,\mathrm{bar}$, $f_2=0\,\mathrm{kHz}$, $\theta=0$ and $R_E=10\,\mathrm{\mu m}$). The first frequency component $f_1$ is varied between $20\,\mathrm{kHz}$ and $1\,\mathrm{MHz}$ with a resolution $N$ distributed logarithmically. Since the amplitude of the second frequency component is zero, the system is driven only by a single frequency. According to the dimensionless form of the time coordinate given by Eq.\,\eqref{dimensionless_time}, the state space is periodic in time with period $\tau_p=1$. At each frequency value, the maximum value of the dimensionless bubble radius $y_1^{max}$ is extracted from the converged solution. The first $1024$ integration phases are regarded as transients and discarded, and the values of $y_1^{max}$ is determined from the subsequent $64$ integration phases. A single integration phase means solving the Keller--Miksis equation over the time domain $\tau=[0, 1]$. The plot of $y_1^{max}$ as a function of $f_1$ with a resolution of $N=46080$ is depicted in Fig.\,\ref{Fig:FrequencyResponseCurve}. Such a function is called amplification diagram or frequency response curve. The objective of this section is to compare the runtimes of the computations of such curves with different resolutions $N$.

\begin{figure}[ht!]    
	\centering
		\includegraphics[width=8.6cm]{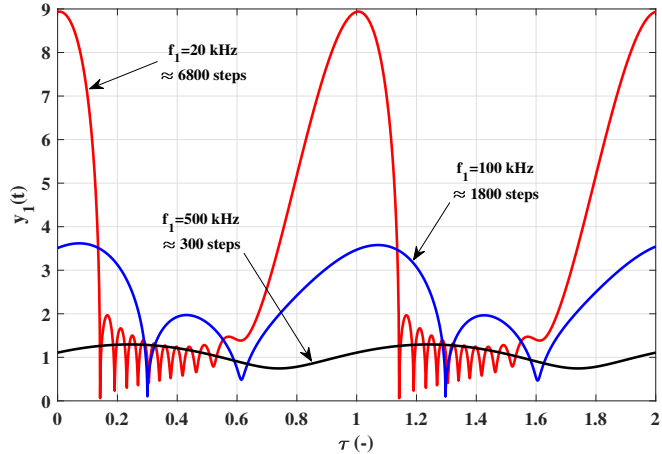}
	\caption{Typical time series of bubble radii at frequency values $f_1=20\,\mathrm{kHz}$ (red), $100\,\mathrm{kHz}$ (blue) and $500\,\mathrm{kHz}$ (black). With the Runge--Kutta--Dormand--Prince solver with $10^{-10}$ absolute and relative tolerances, the required time steps are approximately $6800$, $1800$ and $300$.}
	\label{Fig:BubbleRadiusCurves}
\end{figure}

The source of the challenge of this problem is the qualitatively different dynamics of the bubbles at different frequency values. This is demonstrated in Fig.\,\ref{Fig:BubbleRadiusCurves} where the time series of the dimensionless bubble radius of $2$ integration phases (after the transients) are plotted for $f_1=20\,\mathrm{kHz}$ (red), $100\,\mathrm{kHz}$ (blue) and $500\,\mathrm{kHz}$ (black). The bubble dynamics at $20\,\mathrm{kHz}$ has a slow expansion phase followed by a very rapid contraction (bubble collapse) with some ``afterbounces''. Clearly, the solution has orders of magnitude difference in its time scales; thus, the application of adaptive solvers is mandatory. With the Runge--Kutta--Dormand--Prince scheme with $10^{-10}$ absolute and relative tolerances, the maximum and the minimum time steps are $1.2 \cdot 10^{-3}$ and $2.2 \cdot 10^{-9}$, respectively. With increasing frequency, the dynamics become much smoother. The required time steps to keep the prescribed tolerance (given above) are approximately $6800$ ($20\,\mathrm{kHz}$), $1800$ ($100\,\mathrm{kHz}$) and $300$ ($500\,\mathrm{kHz}$).

In case of such systems, it is extremely important to be able to separate the time coordinate of each instance and solve them asynchronously. Otherwise, in packed systems, the time step will always be determined by the smallest required time step in an ensemble. Therefore, these instances might slow each other down resulting even in orders of magnitude larger number of steps than required. For CPUs, this is a minor issue as with VCL library without unrolling, only a maximum of $4$ Keller--Miksis equations are packed together. However, some program packages can handle ensemble simulations on GPUs only by packing the whole ensemble of oscillators into a single monolithic ODE function. In these cases, one can expect quite poor performance.

\subsubsection{Performance curves on CPUs and GPUs} \label{Sec:KellerMikisPerformanceCurvesCPUGPU}

\begin{figure}[ht!]    
	\centering
		\includegraphics[width=8.6cm]{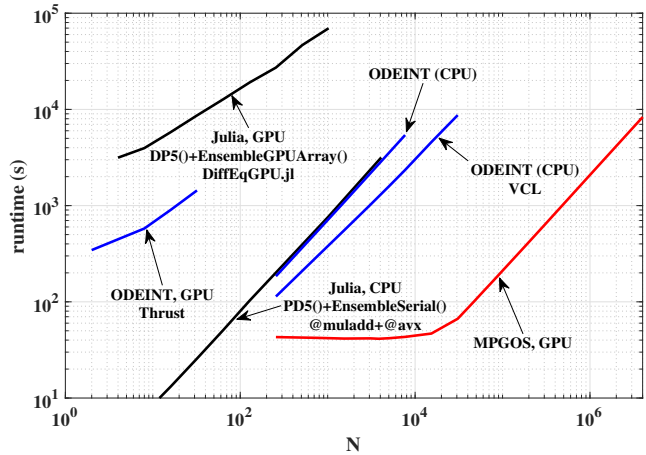}
	\caption{Performance curves of the Keller--Miksis equation; that is, the runtime of the problem defined in this section as a function of the ensemble size $N$. Red curve: MPGOS; blue curves: ODEINT; black curves: Julia (DifferentialEquations.jl).}
	\label{Fig:KellerMiksisPerformanceCurve}
\end{figure}

The performance characteristics are summarised in Fig.\,\ref{Fig:KellerMiksisPerformanceCurve}, where the total runtime of the complete problem is plotted as a function of the resolution $N$ of the frequency range. The runtime involves the $1024$ transient integration phases and an additional $64$ to determine $y_1^{max}$. The colour coding is the same as in the case of the previous figures: the red, blue and black curves are related to MPGOS, ODEINT and Julia, respectively. To simplify the discussion, hand-tuned versions and results taken from other publications are omitted here; and curves corresponding to both the CPU and GPU computations are examined together. As usual, the simulations are performed on an Intel Core i7-4820K CPU (single core, $30.4$ GFLOPS) and on an Nvidia Titan Black GPU ($1707$ GFLOPS). The implementations and the source codes can be found in the GitHub repository \cite{Keller_Miksis_RK45_GitHub}.

In general, the precise location of $y_1^{max}$ is no goal, it is determined as the maximum of the series of points of $y_1$ given at the locations of the natural time steps of the adaptive solvers. Therefore, the overhead of the root finding of an event handling algorithm is excluded. In addition, whenever it was feasible, the determination of $y_1^{max}$ is done without the production of dense output. This can be done via special functions called after every successful time step, in which only a single parameter is continuously updated. These functions are called \textit{ActionAfterSuccessfulTimeStep} (MPGOS), \textit{observer} (ODEINT) and \textit{CallbackFunctions} (Julia). In the case of using GPUs, it is extremely important to avoid large data transfer through the slow PCI-E bus. In addition, large global memory requirements per thread can significantly reduce the number of the residing threads in a single run, which might decrease the utilisation of the GPU and the latency hiding capabilities, thus performance. In the CPU versions, there are no observable differences in the runtimes between employing the special functions or the dense output. In MPGOS and ODEINT, the Runge--Kutta--Cash--Karp algorithm, while in Julia, the Runge--Kutta--Dormand--Prince scheme is applied. Both are $5^{th}$ order solvers with $4^{th}$ order embedded error estimation. In all the cases, both the relative and absolute tolerances of the integrators are $10^{-10}$.

The fastest implementation is provided by MPGOS (GPU hardware). The feature of the performance curve is very similar to the one of the Lorenz system. There is an initial plateau in the runtimes where the GPU is not fully utilised. This phase is followed by the linear characteristics (doubling the work doubles the runtime). The implementation of MPGOS is very efficient. It solves the ensemble of the Keller--Miksis equations asynchronously; therefore, they do not slow each other down at the collapses, see the related discussion in Sec.\,\ref{Sec:KellerMiksisEquation}. Naturally, thread divergence is presented due to the adaptive solvers and the possible large differences in the required number of time steps at different parameter values. However, the asynchronous approach is a viable option here. Compare for example with the GPU versions of ODEINT and Julia, where the ensemble of Keller--Miksis equations can only be solved as a single large ODE system. These implementations are orders of magnitude slower compared to their CPU versions. Thus they are not discussed further. In addition, MPGOS does not use these output. Only a single parameter is updated after every successful time step to obtain $y_1^{max}$.

MPGOS is approximately $130$ times faster than the CPU-ODEINT code used with VCL vector class library (lower blue curve in Fig.\,\ref{Fig:KellerMiksisPerformanceCurve}), where the unroll technique is not employed for minimising the slowdown. Keep in mind that the ratio of the peak double-precision floating-point performance between the hardware is $56$ indicating that MPGOS harnesses the processing power of the hardware better. One reason is the underutilisation of the vector capabilities of the CPU by ODEINT. The VCL version is only $2.3$ times faster than ``ordinary'' ODEINT code. The implementation in Julia (CPU version) is only $1.09$ times slower compared to the non-VCL ODEINT solver. That is, Julia provides a highly optimised code apart from the exploitation of vectorisation. In the case of Julia, the @muladd (for FMA) and the @avx macros are used with an unroll factor of $4$ (this is the minimum required systems for the optimal exploitation of vector registers). However, the @avx macro has no measurable effect on the runtimes, in contrast to the case of the Lorenz system where the speed-up via the @avx macro was approximately a factor $2$. Unfortunately, we could not find out the reason of this non-robust behaviour. In the CPU versions, using the dense output to determine $y_1^{max}$ has no measurable difference in the runtimes. The speed-up factors $\eta$ between the different program packages discussed above are summarised in the first line in Tab.\,\ref{Tab:SpeedupFactors}.

\begin{table}[ht!]
\caption{\label{Tab:SpeedupFactors}Summary of the speed-up factors between the different program packages.}
\centering
  \begin{tabular}{ccccc}
		& MPGOS  & MPGOS   & ODEINT   & MPGOS      \\
	  & Julia  & ODEINT  & Julia    & ODEINT VCL \\
	\hline \hline
		$\eta$   & 325     & 299      & 1.09      & 130     \\
		$\eta_W$ & 285     & 287      & 0.99      & 124     \\
		$\eta_E$ & 196-342 & 238-309  & 0.63-1.44 & 103-134 \\
	\hline
  \end{tabular}
\end{table}

\subsubsection{Relationship between the tolerances, accuracy and the amount of work} \label{Sec:KellerMikisTolerancesAccuracy}

In Fig.\,\ref{Fig:KellerMiksisPerformanceCurve}, the performance curves are plotted with both absolute and relative tolerances of $10^{-10}$. This is reasonable as the user can manage the global error of the solution via the desired local truncation error. However, the global error over a longer integration phase can vary between the program packages. The reasons are the different types of numerical algorithms employed (Cash--Karp and Dormand--Prince), and the slightly different approach of handling the error control. Therefore, the issue of the tolerances of the local error, the achieved global error (accuracy) and the amount of function evaluations (work) need to be addressed. \textit{In this section, the values of the absolute and relative tolerances are the same for every simulation; therefore from now on, we shall refer to them only as the tolerance $T$ for simplifying the discussion.}

\begin{figure}[ht!]    
	\centering
		\includegraphics[width=8.6cm]{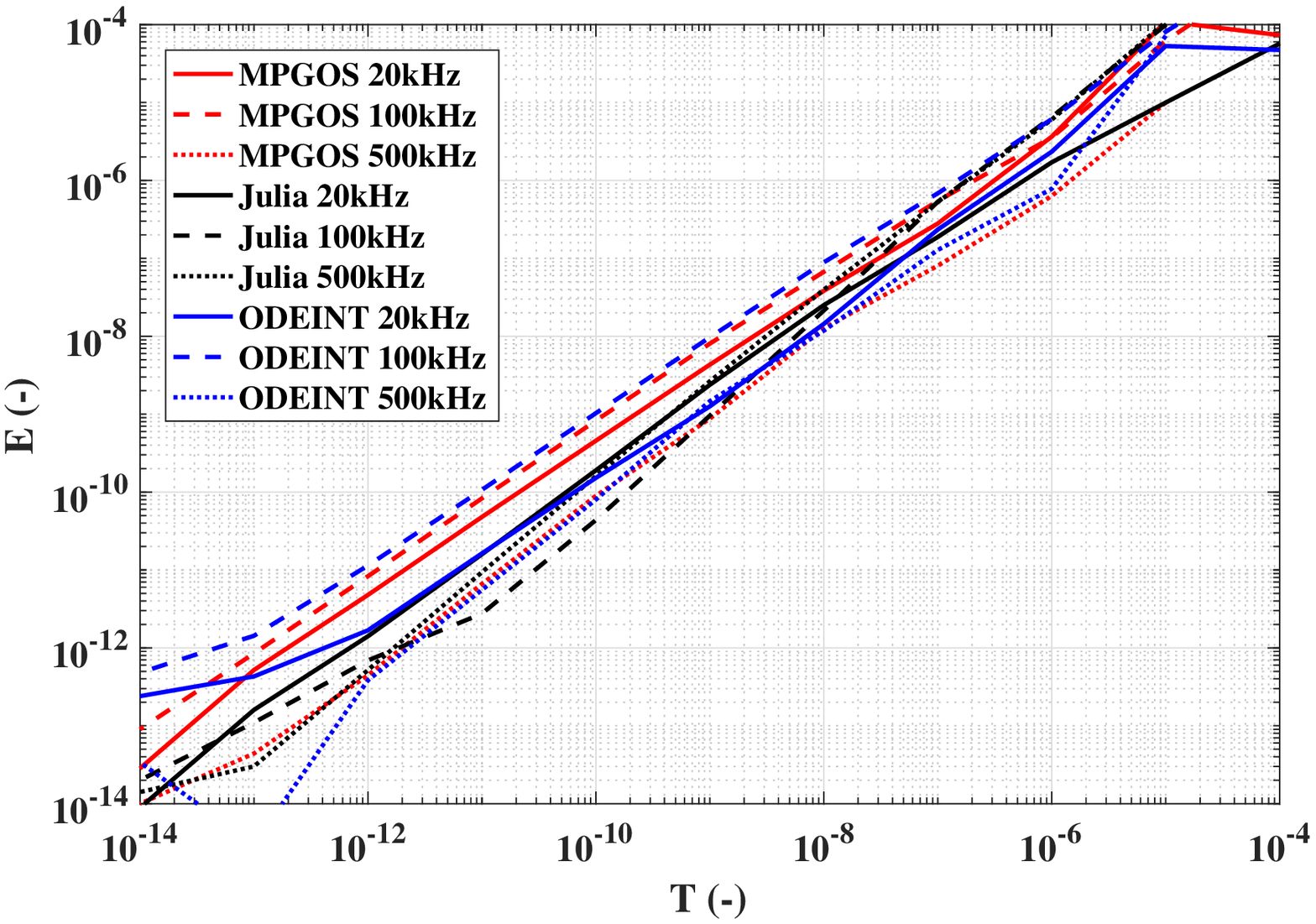}
		\includegraphics[width=8.6cm]{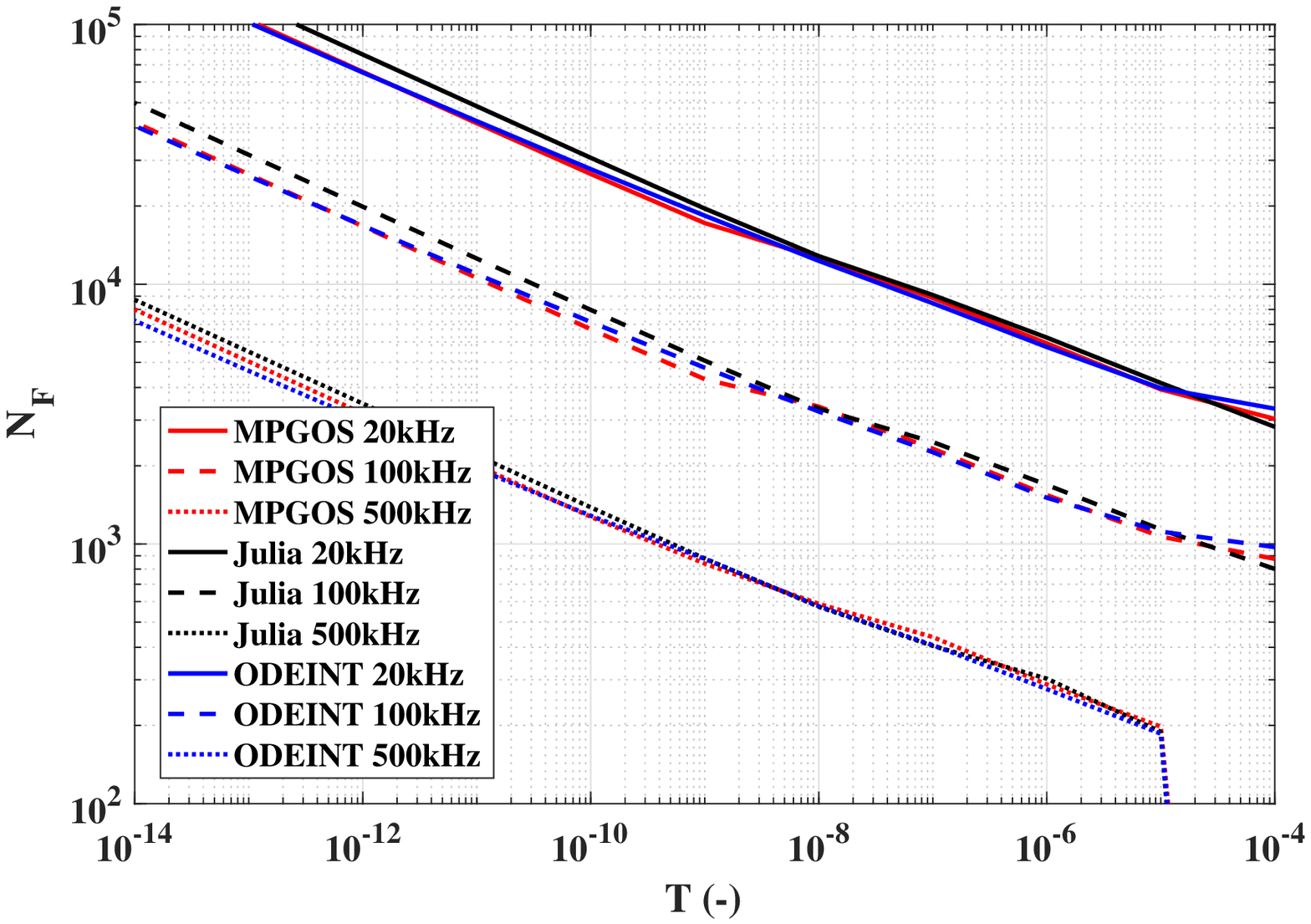}
		\includegraphics[width=8.6cm]{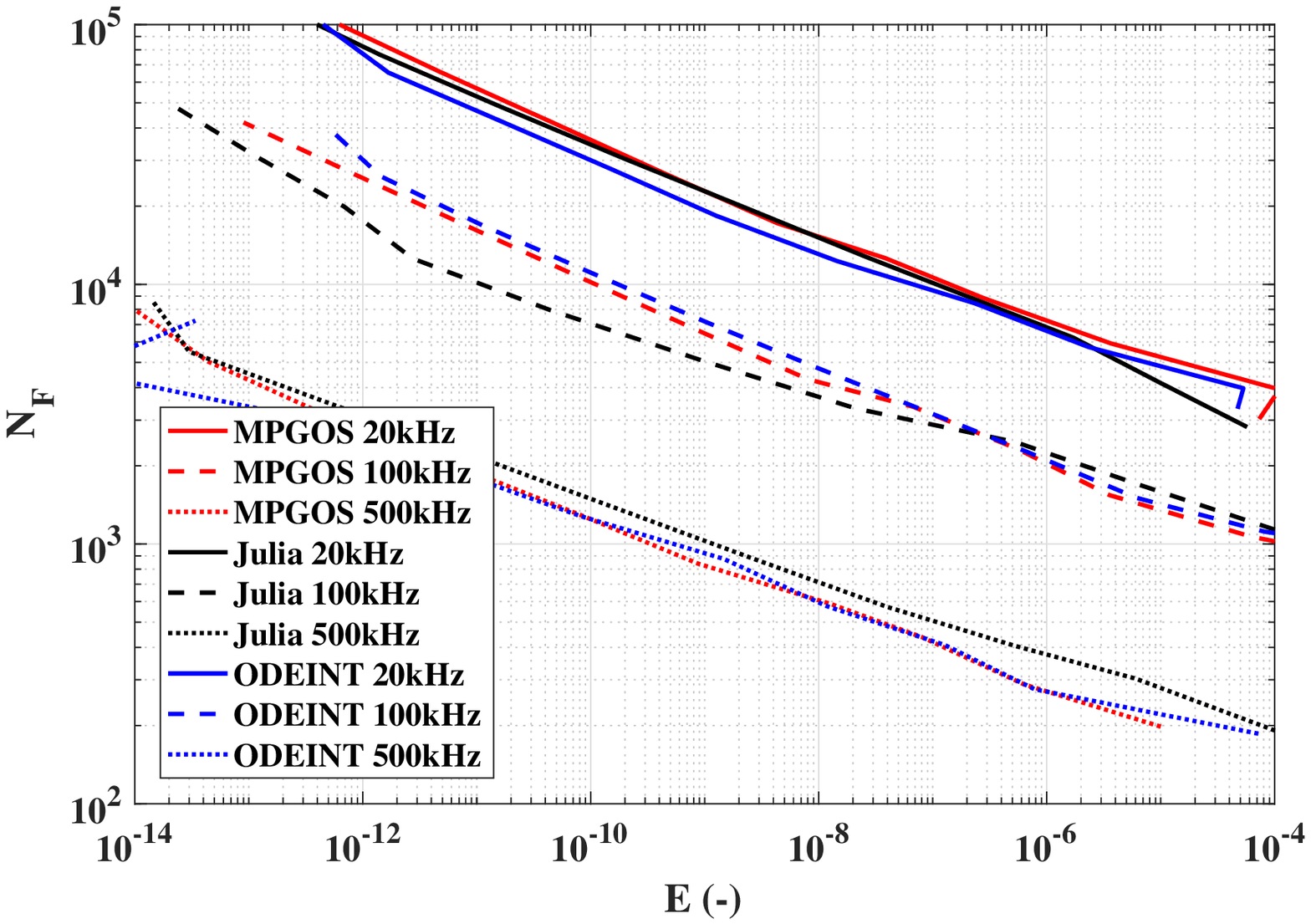}
	\caption{Correlation between the tolerance $T$, global error $E$ and number of function evaluations $N_F$.}
	\label{Fig:ToleranceError}
\end{figure}

One of the major problems with the Keller--Miksis equation is the absence of analytical solutions for calculating the global error precisely. Therefore, numerical experiments are done on a large scale of tolerances (from $T=10^{-4}$ to $10^{-15}$), where the solution with the most stringent tolerance ($10^{-15}$) is considered as ``exact''. This approach is feasible as the ``exact'' solution is accurate up more than $4$ decimal places compared to the solutions with $T=10^{-10}$ (our main interest here). In addition, the long term solution can be chaotic, that prevents any direct comparison of the global error on a large integration time domain. Thus, the integration time domain is restricted to $\tau=[0, 2]$ (two excitation periods) to minimise the effect of chaotic solutions, and the values of the first components of the state space at the integration phase $y_1(2)$ are compared. In this way, the absolute error can be defined as $E=|y_1(2)-y_1^E(2)|$, where $y_1^E(2)$ corresponds to the solution with tolerance $T=10^{-15}$ (the ``exact'' solution). Performing detailed analysis on a large number of frequencies is a cumbersome task; thus, only three characteristic frequency values are examined. Namely, $f_1=20\,\mathrm{kHz}$, $100\,\mathrm{kHz}$ and $500\,\mathrm{kHz}$, see again Fig.\,\ref{Fig:BubbleRadiusCurves}. In the case of ODEINT, the VCL library is not used in this investigation to be consistent with the Julia code. During the comparison of MPGOS and the VCL version of ODEINT, we assumed a constant performance difference of a factor of $2.3$ according to the results of Fig.\,\ref{Fig:KellerMiksisPerformanceCurve}.

The results of the numerical experiments are summarised in Fig.\,\ref{Fig:ToleranceError}. As usual, the colour code is as follows: red is MPGOS, blue is ODEINT and black is Julia. The solid, dashed and the dotted lines are related to $f_1=20\,\mathrm{kHz}$, $100\,\mathrm{kHz}$ and $500\,\mathrm{kHz}$, respectively. The upper panel shows the error $E$ as a function of the tolerance $T$, which is important in verifying the convergence of the solutions by applying more stringent tolerance. However, this graph alone is not suitable for judging the real performance of a solver as the lower global error might also be the result of a higher number of steps.

Thus, the number of the total function evaluations $N_F$ (a measure of the total amount of work) has to be included in the analysis. In the middle panel of Fig.\,\ref{Fig:ToleranceError}, it is plotted as a function of the tolerance. Although Julia always needs slightly more function evaluations (Dormand--Prince) compared to the other program packages (Cash--Karp), for a fixed frequency value, $N_F$ is approximately the same for all cases. Note how the curves having the same style run close to each other. For tolerance $T=10^{-10}$, the data is presented in Tab.\,\ref{Tab:NumberFunctionEvaluation}, where $N_F^M$, $N_F^J$ and $N_F^O$ are the total number of the function evaluations of MPGOS, Julia and ODEINT, respectively. The results of MPGOS are considered as the baseline, and Tab.\,\ref{Tab:NumberFunctionEvaluation} also shows the ratio with Julia ($\approx 14\%$ average increase) and ODEINT ($\approx 4\%$ average increase). Accordingly, slightly modified speed-up factors $\eta_W$ can be defined with the average ratio of the function evaluations to indicate the speed-up for the same amount of work. These are listed in the second line of Tab.\,\ref{Tab:SpeedupFactors}. For instance, between MPGOS and Julia, it is calculated with the following expression
\begin{equation}
\eta_W = \eta / 1.14.
\end{equation}
Note that in terms of $\eta_W$, the $9\%$ performance difference between ODEINT and Julia disappears. Since, Julia does approximately $9\%$ more work (function evaluation).

\begin{table}[ht!]
\caption{\label{Tab:NumberFunctionEvaluation}Number of function evaluations $N_F^M$ (MPGOS), $N_F^J$ (Julia) and $N_F^O$ (ODEINT) for different frequencies using a tolerance of $T=10^{-10}$. The last line shows the average ratio of the corresponding function evaluations.}
\centering
  \begin{tabular}{c|c|cc|cc}
		f    & $N_F^M$ & $N_F^J$ & $N_F^M/N_F^J$ & $N_F^O$ & $N_F^M/N_F^O$ \\
	\hline \hline
		20   & 26580   & 30675   & 1.154         & 27744   & 1.044 \\
		100  & 6714    & 7965    & 1.186         & 7152    & 1.065 \\
		500  & 1272    & 1383    & 1.084         & 1282    & 1.008 \\
	\hline
	\multicolumn{1}{c}{avg.} & \multicolumn{1}{c}{} & \multicolumn{1}{c}{} & \multicolumn{1}{c}{$\approx 1.14 $} & \multicolumn{1}{c}{} & \multicolumn{1}{c}{$\approx 1.04 $} \\
  \end{tabular}
\end{table}

Finally, the speed-up factors can further be adjusted to determine the performance differences for a prescribed global error. In the bottom panel of Fig.\,\ref{Fig:ToleranceError}, the number of the function evaluations $N_F$ is plotted as a function of the global error $E$. Using this diagram, for $E=10^{-10}$, the data are presented in Tab.\,\ref{Tab:NumberFunctionEvaluationGlobalError}. For simplicity, from now on, the notations $N_F^M$, $N_F^J$ and $N_F^O$ are related to a global error of $E=10^{-10}$ instead of to the tolerance of $T=10^{-10}$. The number of function evaluations in Tab.\,\ref{Tab:NumberFunctionEvaluationGlobalError} show much larger deviations between the different program packages compared to the values in Tab.\,\ref{Tab:NumberFunctionEvaluation}. For instance, Julia needs more than $30\%$ fewer function evaluations than MPGOS to keep the global error below $E=10^{-10}$ at frequency value $f=100\,\mathrm{kHz}$. However, the trend is completely the opposite in the case of $f=20\,\mathrm{kHz}$; it needs approximately $20\%$ more function evaluation. Therefore, instead of determining an average modification factor, we present a range of speed-up factors according to the worst and best-case scenarios. The speed-up factors in terms of fixed global error are denoted by $\eta_E$, and between MPGOS and Julia its range is calculated as follows
\begin{equation}
	\begin{split}
		\eta_E^{min} = 0.69 \cdot \eta_W, \\
		\eta_E^{max} = 1.20 \cdot \eta_W.
	\end{split}
\end{equation}
The ranges of the adjusted speed-up factors are printed in the last line of Tab.\,\ref{Tab:SpeedupFactors}. Despite the large differences in the minimum and maximum values, MPGOS is the only viable option for GPUs (also with good performance). In addition, in terms of $\eta_E$, ODEINT and Julia have approximately the same performance on average, although there are large differences between the minimum and maximum values of $\eta_E$.

\begin{table}[ht!]
\caption{\label{Tab:NumberFunctionEvaluationGlobalError}Number of function evaluations $N_F^M$ (MPGOS), $N_F^J$ (Julia) and $N_F^O$ (ODEINT) for different frequencies using a global error of $E=10^{-10}$.}
\centering
  \begin{tabular}{c|c|cc|cc}
		f    & $N_F^M$ & $N_F^J$ & $N_F^M/N_F^J$ & $N_F^O$ & $N_F^M/N_F^O$ \\
	\hline \hline
		20   & 36000   & 34500   & 0.96          & 30000   & 0.83 \\
		100  & 10200   & 7050    & 0.69          & 11100   & 1.08 \\
		500  & 1246    & 1492    & 1.20          & 1247    & 1.00 \\
	\hline
  \end{tabular}
\end{table}

\subsection{A system exhibiting impact dynamics} \label{Sec:ImpactDynamics}

The last test case is a model that describes the dynamics of a pressure relief valve. The numerical difficulty of this problem is the possible non-smooth impacting behaviour. The dimensionless governing equations are taken from \cite{Hos2012a} and can be written as
\begin{align}
\dot{y}_1  &= y_2, \label{pressure_relief_valve_1} \\
\dot{y}_2  &= -\kappa y_2 - (y_1+\delta) + y_3, \label{pressure_relief_valve_2} \\
\dot{y}_3  &= \beta ( q - y_1 \sqrt{y_3} ), \label{pressure_relief_valve_3}
\end{align}
where $y_1$ and $y_2$ are the displacement and the velocity of the valve body, respectively. The pressure relief valve is attached to a reservoir chamber in which the dimensionless pressure is $y_3$. The control parameter of the system is the dimensionless flow rate $q$ varied between $0.2$ and $10$ with a resolution of $N$ (uniform distribution). The rest of the parameters are kept constant: $\kappa=1.25$ is the damping coefficient, $\delta=10$ is the precompression parameter, $\beta=20$ is the compressibility parameter.

In Eqs.\,\eqref{pressure_relief_valve_1}-\eqref{pressure_relief_valve_3}, the zero value of the displacement ($y_1=0$) means that the valve body is in contact with the seat of the valve. If the velocity of the valve body $y_2$ has a non-zero, negative value at this point, the following impact law is applied:
\begin{align}
y_1^+  &= y_1^- = 0, \label{impact_law_1} \\
y_2^+  &= -r y_2^-, \label{impact_law_2} \\
y_3^+  &= y_3^- \label{impact_law_3}
\end{align}
That is, the velocity of the valve body is reversed by the Newtonian coefficient of restitution $r=0.8$ that approximates the loss of energy of the impact.

\begin{figure}[ht!]    
	\centering
		\includegraphics[width=8.6cm]{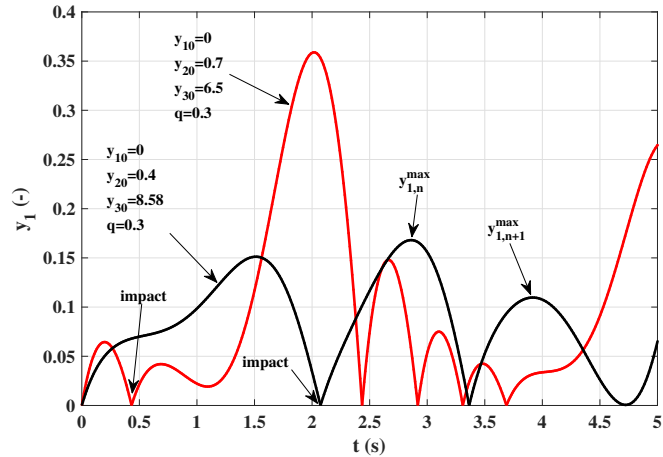}
	\caption{Typical impacting solutions of the pressure relief valve model described by Eqs.\,\eqref{impact_law_1}-\eqref{impact_law_3} employing different initial conditions.}
	\label{Fig:ImpactBehaviour}
\end{figure}

Figure\,\ref{Fig:ImpactBehaviour} shows two examples of impacting solutions at different initial conditions; the control parameter is $q=0.3$ in both cases. The first numerical challenge is that the different instances of the ODE system can have an impact at different time instances. It is impossible to handle this situation properly when a large number of instances are packed into a single monolithic system as the precise detection of the location of the impact has to be serialised. That is, if a single instance is impacting, all the other instances are blocked from progressing further during the precise impact detection. The necessary control flow operations must be serialised as well. The main reason is that the SIMD lanes in case of vector registers (CPUs) and the threads in a warp (GPU) have to perform the same instruction but on multiple data. For CPUs, it is a minor drawback as usually only a few instances are packed together ($4$ is enough for Ivy Bridge). Thus, the performance drop can be acceptable. However, in the case of GPUs, where thousands of threads have to work together and block each other, the performance loss can be very severe.

Another issue comes from the autonomous nature of the system. For a feasible investigation of the periodic orbits, a suitable definition of the Poincar\'e section is mandatory. In case of the pressure relief valve model introduced here, a possible option is the local maximum of the displacement. Therefore, a single integration phase means the integration from a local maximum $y_{1,n}^{max}$ to the next local maximum $y_{1,n+1}^{max}$, see again Fig.\,\ref{Fig:ImpactBehaviour}. This technique needs another event detection procedure for the local maxima (besides the detection of the impact). The consequence of such a definition of the Poincar\'e section is a difference between the integration time domains of each ODE system instances. Thus, in the case of packed ODEs, it is impossible to stop the integration phases properly. This is the reason why performance curves are not presented for Julia and ODEINT using GPUs in Sec.\,\ref{Sec:ImpactDynamicsPerformanceCurvesCPUGPU}.

Since MPGOS solves all the instances of the ODE system asynchronously, these problems are not serious issues. Naturally, a small amount of thread divergence occurs due to the employed adaptive solver, the detection of the events and the application of the impact law for individual threads. However, conditionals are automatically masked in a warp; thus, it needs no further control logic by the user. In contrast, using the vector registers of a CPU, the divergent execution path in an SIMD lane must be handled by the programmer separately producing some amount overhead. Furthermore, the divergent part in an event detection in GPUs is restricted only to the determination of a corrected time step and the application of the very simple impact law. These operations need a fare less number of instructions compared to a complete time step. This makes MPGOS a very efficient solver. For the details, the interested reader is referred again to our preliminary publication \cite{Hegedus2020b} or to the manual of the program package \cite{Hegedus2019a}.

\begin{figure}[ht!]    
	\centering
		\includegraphics[width=8.6cm]{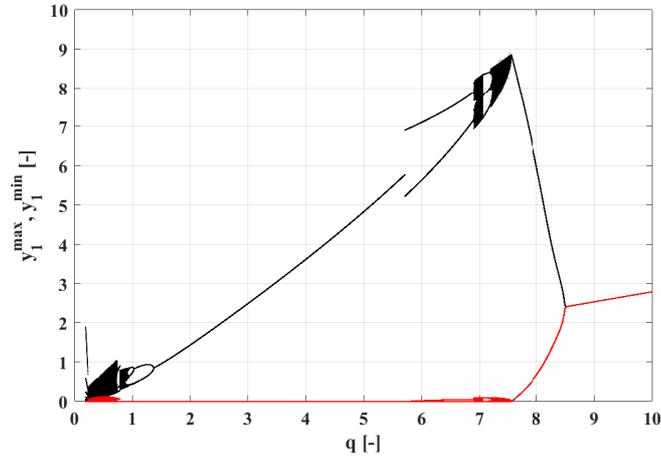}
	\caption{Bifurcation diagram where the maximum (black) and the minimum (red) values of the valve position $y_1$ are plotted as a function of the dimensionless flow rate $q$.}
	\label{Fig:PressureReliefValve}
\end{figure}

The objective of this example is the study of the numerical behaviour of the different program packages instead of providing detailed physical investigation and interpretation of the results. In order to achieve this, a bifurcation diagram is generated, where the maximum and minimum displacement of the valve body ($y_1^{max}$ and $y_1^{min}$) through $32$ integration phases are plotted as a function of the control parameter $q$ (dimensionless flow rate). Note that the values of $y_1^{max}$ are the points of the Poincar\'e section. The values of $y_1^{min}$ are registered only to recognise impacting solutions ($y_1^{min}=0$). Again, the control parameter $q$ is varied between $0.2$ and $10$, and the resolution $N$ is the parameter for the performance characteristic curve. In each simulation, the first $1024$ iterations are regarded as initial transients and discarded. The data of the next $32$ iterations are recorded and written into a text file. A bifurcation diagram with $N=30720$ is shown in Fig.\,\ref{Fig:PressureReliefValve}. The black and red dots are the values of $y_1^{max}$ and $y_1^{min}$, respectively. It is clear that in a wide range of the control parameter (approximately between $0.2$ and $7.5$), the solutions are impacting. This puts the program packages to the test.

\subsubsection{Performance curves on CPUs and GPUs} \label{Sec:ImpactDynamicsPerformanceCurvesCPUGPU}

\begin{figure}[ht!]    
	\centering
		\includegraphics[width=8.6cm]{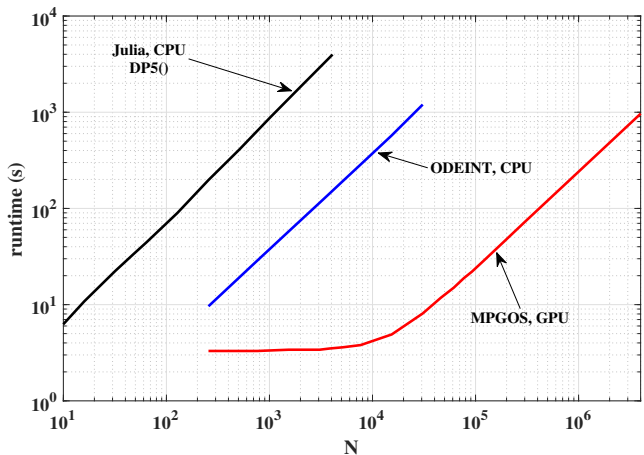}
	\caption{Performance curves of the pressure relief valve model; that is, the runtime of the problem defined in this section as a function of the ensemble size $N$. Red curve: MPGOS; blue curve: ODEINT; black curve: Julia (DifferentialEquations.jl).}
	\label{Fig:PressureReliefValvePerformanceCurve}
\end{figure}

In this section, the performance characteristic curves are discussed shortly, as all the basics are already examined with great detail in the previous sections. They are shown in Fig.\,\ref{Fig:PressureReliefValvePerformanceCurve}, where the total runtime of the complete problem is plotted as a function of the resolution $N$ of the control parameter $q$. The runtime involves the $1024$ transient integration phases and an additional $32$ to determine the values of $y_1^{max}$ and $y_1^{min}$. The colour coding is the same as in the previous cases: the red, blue and black curves are related to MPGOS, ODEINT and Julia, respectively. Due to the aforementioned issues (event handling, impact law and different time domains), only the results of MPGOS are depicted as a sole GPU performance curve. The hardware is the Nvidia Titan Black GPU ($1707$ GFLOPS). Due to the same complications, the CPU versions do not use the vector registers via the AVX instruction set. In our experience, the vectorised versions have no observable benefit in this problem. The employed CPU is the Intel Core i7-4820K CPU (single core, $30.4$ GFLOPS). Similarly to the example of the Keller--Miksis equation, in MPGOS and ODEINT, the Runge--Kutta--Cash--Karp algorithm, while in Julia, the Runge--Kutta--Dormand--Prince scheme is applied. Both the relative and absolute tolerances of the integrators are $10^{-10}$. The absolute tolerance of the impact detection is set to $10^{-6}$.

It should be noted that ODEINT does not support automatic root-finding of special points (events). When a possible event is detected via the \textit{observer}, the user has to implement an own algorithm (we used the bisection method) to find the precise location of the events within a prescribed tolerance. It is important in this case as the global error of the solution is sensitive to the proper application of the impact law.

In the asymptotic (linear) regime, MPGOS is approximately $148$ times faster than ODEINT, which is much higher than the ratio of the peak performance difference of the employed hardwares ($\times 57$). This indicates a very efficient implementation of the problem in MPGOS. Comparing the CPU versions, the ODEINT solver is about $26$ times faster than Julia. This result is surprising as Julia has approximately the same performance as ODEINT in the case of the Keller--Miksis equation (Sec.\,\ref{Sec:KellerMiksisEquation}) and only has slightly larger runtimes for the Lorenz system (Sec.\,\ref{Sec:LorenzSystem}). It implies a suboptimal event handling by Julia. For further details of the source codes and implementations, see the GitHub repository of the problem \cite{PRV_RK45_GitHub}.

\section{Summary and conclusion} \label{Sec:Conclusion}

The main aim of the present study was to provide a detailed performance comparison of different program packages to solve a large number of independent, low-order, non-stiff ordinary differential equation (ODE) systems. Both the CPU and GPU capabilities of the ODE suits were tested. The three candidates were MPGOS, designed for using only GPUs, ODEINT, written in C++ and DifferentialEquations.jl, implemented in Julia. Three models were tested, each having different features and numerical challenges. First, the Lorenz system was investigated employing the classic fourth-order Runge--Kutta solver (fixed time step). This is a standard example in many program packages and textbooks for measuring the performance of a specific solver. Second, the Keller--Miksis equation known in sonochemistry and bubble dynamics was tested. Here, the large time-scale differences presented in a solution pose a real challenge when exploiting the SIMD units of a hardware, or if a single monolithic ODE system is built-up from a multitude of Keller--Miksis equations. The last example was a model describing the dynamics of a pressure relief valve that can exhibit impacting behaviour. In this case, the solvers needed to handle multiple events and the non-smooth nature of the system (impact). In the last two models, adaptive Runge--Kutta algorithms were used (Cash--Karp or Dormand--Prince).

As a general conclusion, MPGOS is superior using GPUs. In many cases, it has orders of magnitude lower runtimes compared to the other program packages. In addition, it is the only package that could handle the non-smooth dynamical system (pressure relief valve). Therefore, it is a perfect tool for parameter studies and non-linear analysis in high dimensional parameter space \cite{Freire2020a,Varga2020a,Marcondes2017a,Gallas2015a,Freire2011a,Freire2011b,deSouza2012a,Medeiros2011a,Medeiros2010a,Medrano2014a,Celestino2014a,Nicolau2019a,Jousseph2018a,Jousseph2016a}. For CPUs, in many cases, ODEINT and Julia have approximately the same performance. Therefore, Julia is a viable option compared to other program packages written in low-level languages. However, Julia has some overhead in case of event handling, and ODEINT is superior in these problems. It must also be noted that ODEINT lacks root finding, and for precise event detection, the users have to write their own algorithm.

Our personal impression about the program packages is as follows. Julia is a high-level programming language and thus specifically designed to be user-friendly allowing quick code development with minimal learning time. Its CPU performance can be considered excellent (except for cases with event handling), but for GPUs, the runtimes show poor efficiency. Nevertheless, it must be stressed that the package has many other features not shown here; for instance, automatic differentiation to calculate the Jacobi for stiff problems, using arbitrary-precision numbers or the support of delay differential equations, to name a few. Moreover, it has excellent support, and the newer versions of the code are expected with performance improvements.

ODEINT is written in the low-level C++ programming style. Due to the exhaustive usage of template metaprogramming, the CPU related codes have excellent performance, and it provides a relatively easy means to fuse the application with other libraries. For instance, with the Vector Class Library for explicit vectorisation. It provides multistep and symplectic integrators, as well as stiff solvers, too. However, the package is far from user-friendly. One needs a very long learning period to be able to build-up applications dissimilar to the provided tutorial examples. The performance in terms of GPU usage is poor.

MPGOS fuses the user-friendliness and high-performance into a single ODE suite. Its interface is easy to use, and the right-hand side of the ODE function can be implemented in a similar way as in the case of Julia or MATLAB. In addition, it also provides low-level means for fine-tuning and handling specialised problems. The main drawback of MPGOS is the lack of additional features as those of Julia and ODEINT; it supports only Runge--Kutta type solvers. However, MPGOS is continuously under development, and the interested user should regularly check its new features in the official website \cite{Hegedus_MPGOS_WebSite}.

\section*{Acknowledgement}
The research reported in this paper and carried out at BME has been supported by the NRDI Fund (TKP2020 IES, Grant No. BME-IE-BIO) based on the charter of bolster issued by the NRDI Office under the auspices of the Ministry for Innovation and Technology. This paper was also supported by the Alexander von Humboldt Foundation (HUN 1162727 HFST-E), by the J\'anos Bolyai Research Scholarship of the Hungarian Academy of Sciences (Ferenc Heged\H us), and by the New National Excellence Program of the Ministry of Human Capacities, project no. \'UNKP-20-1-I-BME-168 (D\'aniel Nagy) and \'UNKP-20-1-I-BME-186 (Lambert Plavecz).


\bibliographystyle{elsarticle-num}
\biboptions{sort&compress}

\bibliography{2020f_ComputPhysCommun}

\end{document}